\pdfoutput=1
\documentclass[12pt,reqno]{amsart}
\usepackage{amsmath,amsfonts,amssymb,amscd,amsthm,amsbsy, color}
\usepackage[pdftex]{graphicx}
\usepackage{appendix}
\usepackage{comment}

\numberwithin{equation}{section}

\newtheorem{theorem}{Theorem}
\newtheorem{meta-thm}[theorem]{Meta-Theorem}

\newtheorem{remark}[theorem]{Remark}
\newtheorem{definition}[theorem]{Definition}

\newcommand\beq[1]{ \begin{equation}\label{#1} }
\newcommand{\eeq}{ \end{equation} }

\newcommand\beqa[1]{ \begin{eqnarray} \label{#1}}
\newcommand{\eeqa}{ \end{eqnarray} }
\newcommand{\beqano}{ \begin{eqnarray*} }
\newcommand{\eeqano}{ \end{eqnarray*} }
\newcommand\equ[1]{{\rm (\ref{#1})}}

\def\A{{\mathcal A}}

\def\H{{\mathcal H}}

\def\F{{\mathcal F}}
\def\G{{\mathcal G}}
\def\H{{\mathcal H}}

\def\R{{\mathcal R}}

\def\integer{{\mathbb Z}}

\begin{document}

\title[Dynamical models and the onset of chaos in space debris]
{Dynamical models and the onset of chaos in space debris}

\author[A. Celletti]{Alessandra Celletti}
\address{
Department of Mathematics, University of Rome Tor Vergata, Via della Ricerca Scientifica 1,
00133 Rome (Italy)}
\email{celletti@mat.uniroma2.it}

\author[C. Efthymiopoulos]{Christos Efthymiopoulos}
\address{
Research Center for Astronomy and Applied Mathematics, Academy of Athens
Soranou Efessiou 4, 115 27 Athens (Greece)}
\email{cefthim@academyofathens.gr}

\author[F. Gachet]{Fabien Gachet}
\address{
Department of Mathematics, University of Rome Tor Vergata, Via della Ricerca Scientifica 1,
00133 Rome (Italy)}
\email{gachet@mat.uniroma2.it}

\author[C. Gale\c s]{C\u at\u alin Gale\c s}
\address{
Department of Mathematics, Al. I. Cuza University, Bd. Carol I 11,
700506 Ia\c si (Romania)}
\email{cgales@uaic.ro}

\author[G. Pucacco]{Giuseppe Pucacco}
\address{
Department of Physics, University of Rome Tor Vergata, Via della Ricerca Scientifica 1,
00133 Rome (Italy)}
\email{pucacco@roma2.infn.it}


\baselineskip=12pt              




\begin{abstract}
The increasing threat raised by space debris led to the
development of different mathematical models and approaches to
investigate the dynamics of small particles orbiting around the
Earth. Such models and methods strongly depend on the altitude of
the objects above Earth's surface, since the strength of the
different forces acting on an Earth orbiting object (geopotential,
atmospheric drag, lunar and solar attractions, solar radiation
pressure, etc.) varies with the altitude of the debris.

In this review, our focus is on presenting different
analytical and numerical approaches employed in modern studies of
the space debris problem. We start by considering a model
including the geopotential, solar and lunar gravitational forces
and the solar radiation pressure. We summarize the equations
of motion using different formalisms: Cartesian coordinates,
Hamiltonian formulation using Delaunay and epicyclic variables,
Milankovitch elements. Some of these methods lead in a
straightforward way to the analysis of resonant motions. In
particular, we review results found recently about the
dynamics near tesseral, secular and semi-secular resonances.

As an application of the above methods, we proceed to analyze
a timely subject namely the possible causes for the onset of
chaos in space debris dynamics. Precisely, we discuss the
phenomenon of overlapping of resonances, the effect of a large
area-to-mass ratio, the influence of lunisolar secular resonances.

We conclude with a short discussion about the effect of the
dissipation due to the atmospheric drag and we provide a list of
minor effects, which could influence the dynamics of space debris.
\end{abstract}


\maketitle



\section{Introduction}\label{intro}

%
%
%
%
%

Space activity around the Earth underwent a phenomenal growth in the last decades with very rich benefits for public and private companies,
and ordinary people. Our planet is now surrounded by a huge population of satellites located in all orbital zones with the most
diverse tasks. At the end of their lifetime in the re-entry phase or during breakup
events, fragments are generated and can be injected in different orbits.
Moreover, accidents due to impacts with natural bodies or even collisions between spacecraft greatly contribute to produce whole swarms of new orbiting objects. The proliferation of this crowd of \sl space debris \rm is now realized to be a serious threat to current and future missions,
and a concrete source of risk for man operated spacecraft (\cite{X1,X2}).

Space agencies have seriously considered the extent of the danger and started \sl space situational awareness \rm programs to investigate all sources of hazard both for Earth and its orbiting environments. Among these programs, monitoring and cataloguing the space debris population is now an ongoing activity with an ever enriching database. On the theoretical side, it is of paramount importance to understand the dynamical evolution of this population in order to forecast the most relevant issues and program mitigation strategies. New end-of-life procedures and deployment techniques are currently under development.

Our purpose in the present review is to make a summary of the most important
analytical approximations which have been developed so far in the literature, in order to describe
the dynamical evolution of Earth-orbiting space debris and artificial satellites.
In recent times, several works have been devoted to the modeling of space debris
(see, e.g., \cite{CGmajor,ElyHowell,VLA,VLD} and references therein)
and to the applications of tools from dynamical systems theory to the study of the regular
or chaotic behavior of space debris (see, e.g., \cite{CGminor,Gachet2016,Rosengren2013,VDLC}).
Different approaches have been used in the literature, for example the integration of Cartesian equations,
the use of action-angle Delaunay variables, the formulation of the model through Milankovitch or epicyclic variables.
Each approach has specific advantages and possible drawbacks.
This paper aims to provide a systematic exposition of the main analytical approaches to describe the
dynamics of space debris. We also review the occurrence of different kinds of resonant motions and we
describe possible mechanisms for the onset of a chaotic behavior.

Our review of the dynamics is based on a case classification, taking into account the most general structural  
and operating features characterizing an object at altitudes
varying from the lower atmosphere up to the geostationary orbit. 
The above analysis includes objects characterized by the most general
structural features and operating from the lower atmosphere up to the geostationary orbit.
Our analysis includes the most relevant perturbations of the underlying Keplerian dynamics, namely the gravitational influence of the
geopotential, the attraction of Sun and Moon, the effect of the solar radiation pressure (hereafter SRP) and the atmospheric drag. An appropriate balance of these sources is necessary according to the orbital regions and the physical properties of the orbiting body,
so that each dynamical model is devised by including only non-negligible contributions. This procedure allows us to capture the most relevant aspects of the dynamical evolution with
simple analytical tools. Using such models as starting points,
more refined predictions can be obtained in a further stage with  dedicated numerical integrations.

\vskip.1in

When studying space debris one typically distinguishes between three main regions, referred to as
Low--Earth--Orbits (hereafter LEO), Medium--Earth--Orbits (hereafter MEO) and geosynchronous--Earth--Orbits
(hereafter GEO), which are briefly introduced as follows.

LEO is the region between 90 and 2\,000 $km$, which is affected (in order of importance) by the gravitational attraction
of the Earth, air drag, Earth's oblateness, the attraction of the Moon, the influence of the Sun and the SRP.

MEO is the region between 2\,000 and 30\,000 $km$, which is affected (again in order of importance) by the gravitational attraction
of the Earth, Earth's quadrupole $J_2$, followed by the actions of $J_{22}$, Moon and Sun (with an ordering of these three effects
varying according to the altitude) and SRP. In particular, GPS satellites are located at about 26\,560 $km$
from Earth's center, they orbit with a velocity of about 3.9 $km/sec$ and a
period of $12^h$ (sidereal time), which corresponds to a 2:1 resonance.

GEO is the region at distances larger than 30\,000 $km$, which is affected, in order, by the gravitational attraction
of the Earth, Earth's quadrupole, the attraction of Moon and Sun, higher-order terms of the geopotential
and the SRP. In particular, \sl geostationary \rm  orbits are located at about 42\,164 $km$ from Earth's center with a
period of $24^h$ which corresponds to the 1:1 \sl synchronous \rm resonance. In GEO, any decay mechanism due to atmospheric friction is absent; each satellite in
a geostationary orbit is assigned
to an orbital slot of about $0.1^{\circ}$ of width in longitude. Any debris created in this region will stay there almost forever.

\vskip.1in

The hierarchies of forces listed above determine the relative magnitude of the perturbations and henceforth the models devised to investigate their effects on the orbital elements. The geopotential
and the lunisolar attraction induce long--term and secular variations in some orbital elements, precisely the eccentricity $e$,
the inclination $i$, the argument of perigee $\omega$ and the longitude of the ascending node $\Omega$.
For example, the lunisolar secular resonances cause long--period changes of the eccentricity and can modify the configuration of the navigation
constellations (\cite{aR15,aR08}).
The solar radiation pressure is due
to the absorption and reflection of photons by the surface of the body and depends on the area--to--mass ratio $A/ m$ and on the orientation of the reflecting surface with respect to the direction of the Sun. In the models presented here, we restrict to the case of the
\sl cannonball \rm approximation, namely the simplifying assumption with the surface always perpendicular to Sunlight. This very simple choice has the advantage of including SRP in the set of Hamiltonian perturbations. For $0.01\leq A/ m\leq 0.1$ $m^2/ kg$, the effect of the radiation pressure is mainly a long periodic change of
$e$ and $\omega$ (see, e.g., \cite{Rosengren2013,VLA,VDLC}).

Atmospheric drag is a dissipative effect and therefore one cannot use
Hamiltonian formulations. However, in the LEO region, where friction forces are the dominant perturbation and are quite well modelled,
semi-analytical treatments are possible and the main effects are easily described (compare with \cite{CGleo}).
For a space debris with area--to--mass ratio of the order of 0.01 $m^2/ kg$, the orbital lifetime in a circular orbit with radius of about 250 $km$
is less than 20 days. At 400 $km$ the lifetime is less than 200 days,
while at 800 $km$ it can reach hundreds of years. However, atmospheric models are gradually less accurate in the transition
between LEO and MEO orbits and the corresponding dynamical effects are therefore poorly described.

\vskip.1in

As an indicative list of references for past numerical studies of the
motion in the GEO domain see \cite{VS, FKZ, KH, JAH, J2012, PA, RS}. We emphasize that the analytical
techniques discussed below, besides recovering most of the results found
in numerical studies, offer valuable insight into the interpretation of
the results.

In the Hamiltonian framework, the natural approach is based on canonical perturbation theory. By using appropriate
sets of canonical variables, the geopotential, the lunisolar attraction and the SRP can be expressed in forms suitable for the series expansions needed in perturbation methods. However, for each orbital region with its proper hierarchy of perturbations,
it is useful to select the main contributions and perform a proper ordering of terms, to better grasp the dominant dynamical behaviour (see, e.g.,
\cite{Gachet2016}). This objective is pursued by averaging or by applying more general canonical transformations leading to \sl normal form \rm Hamiltonians.

\vskip.1in

We stress the fundamental fact that different, possibly complementary,
approaches are very useful in order to construct the simplest model to describe each dominant effect.
For this reason, we devote the first part of this paper to review different formulations which are shortly
described below.

\begin{enumerate}
    \item[(i)] The Cartesian setting (see Section~\ref{sec:model}) has a special role in numerical
    simulations, since one can propagate all effects. It often provides a way to test the accuracy of other methods,
especially when based on truncated series expansions of the potential term.
    \item[(ii)] A Hamiltonian approach using Delaunay variables (see Section~\ref{sec:Delaunay})
    leads naturally to the determination of the existence, location and stability of resonances,
    which are of paramount importance for the analysis of the long-term evolution. This approach allows
    also to distinguish between different types of resonances, e.g. tesseral, secular,
    semi-secular (see Section~\ref{sec:resonances}).
    \item[(iii)] A formulation of the problem using Milankovitch elements (see Section~\ref{sec:Milankovitch})
    is directly linked to two vectorial first integrals of the two-body problem; these elements provide
    useful geometrical insights.
    \item[(iv)] A Hamiltonian approach using epicyclic variables (see Section~\ref{sec:epicyclic})
    allows to implement an easier algebra to compute normal forms.
\end{enumerate}

Overall, we are faced with dynamical systems endowed with generic, namely \sl non-integrable, \rm dynamics. This means that, when the perturbations are small,
a phase-space is mostly regular and filled by quasi-periodic orbits. Normal forms can be used to construct approximate integrals of motion,
which can be considered as effective \sl proper elements \rm for them. Increasing the size of the perturbations and/or in presence of strong resonance conditions, \sl chaotic \rm orbits start to play an important role. The assessment of the relevance of chaotic dynamics is another fundamental issue concerning the fate of debris populations.
In Section~\ref{sec:chaos} we will illustrate several case-studies in which the chaotic evolution of the orbital elements
can be traced due to certain classes of perturbations.

\vskip.1in

The structure of this paper is the following: in Section~\ref{sec:model} we provide the setting for the Earth gravitational potential (geopotential)
and the equations of motion in the standard Cartesian coordinates; in Section~\ref{sec:Delaunay} we recall orbital elements and Delaunay variables,
and use them to express the equations of motion in canonical form, including the time-dependent terms due to the Moon and Sun;
in Section~\ref{sec:resonances} resonances are identified and analysed in each orbital region; in Section~\ref{sec:Milankovitch}
we recall Keplerian elements in vector form (\sl Milankovitch elements\rm); Section~\ref{sec:epicyclic}
is devoted to the analysis of the geostationary secular dynamics performed with the aid of \sl epicyclic variables; \rm
in Section~\ref{sec:chaos} we perform a general study of the chaotic dynamics by evaluating the Fast Lyapunov Indicator to discriminate between regular and
chaotic solutions of the equations of motion; in Section~\ref{sec:drag} the semi-analytical models of drag are revised;
in Section~\ref{sec:minor} a list of small additional effects, neglected in the present study, is given;
we give some conclusions in Section~\ref{sec:conclusions}.
In the Appendix we briefly recall the chaos indicators used in Section~\ref{sec:chaos}.

\section{The model in the Cartesian framework}\label{sec:model}
We consider a small body, say a space debris $S$, subject to the gravitational attraction of the Earth,
including the oblateness of our planet, the influence of the Moon, Sun and solar radiation pressure.
For the moment we disregard any other additional force that might affect the dynamics of the space debris,
like Earth's atmosphere, solar wind, Poynting-Robertson effect, etc.

To introduce the equations of motion, we consider two different frames of reference with the origin in the center of the
Earth: $(i)$ a quasi--inertial frame with unit vectors $\{\mathbf{e}_1,\mathbf{e}_2,\mathbf{e}_3\}$ fixed;
$(ii)$ a synodic frame with unit vectors $\{\mathbf{f}_1,\mathbf{f}_2,\mathbf{f}_3\}$ along axes rotating
with the same angular velocity of the Earth.

To get rid of the rotation of the Earth, we introduce the variable $\theta$, which denotes the sidereal time,
defined by the apparent diurnal motion of the stars. Moreover, let $\mathbf{r}$ be the radius
vector associated to $S$ and let $(x,y,z)$ be the coordinates in the quasi--inertial frame,
while $(X,Y,Z)$ denote the coordinates in the synodic frame:
$$
\mathbf{r}=x \mathbf{e}_1+y\mathbf{e}_2+z\mathbf{e}_3 = X\mathbf{f}_1+Y\mathbf{f}_2+Z\mathbf{f}_3\ .
$$
The relation of the unit vectors within the two frames is
$$
\left(\begin{array}{c}
         \mathbf{f}_1 \\
         \mathbf{f}_2 \\
         \mathbf{f}_3 \\
       \end{array}\right)=
R_3(\theta)\left(\begin{array}{c}
         \mathbf{e}_1 \\
         \mathbf{e}_2 \\
         \mathbf{e}_3 \\
       \end{array}\right)\ ,
$$
where $R_3(\theta)$ is the rotation matrix of angle $\theta$ around the third axis. The relation between the
coordinates in the two frames is given by
\begin{equation}\label{catalin1}
\left(\begin{array}{c}
         x \\
         y \\
         z \\
       \end{array}\right)=
R_3(-\theta)\left(\begin{array}{c}
         X \\
         Y \\
         Z \\
       \end{array}\right)\ .
\end{equation}

\subsection{Equations of motion}\label{sec:eq}
Using the above notation, we can write the equations of motion adding the contributions due to the
Earth, including its oblateness, the solar and lunar attractions, as well as the solar radiation pressure:
\beqa{eq1}
\ddot{\mathbf{r}}=&-&\mathcal{G}\int_{V_E} \rho(\mathbf{r}_p)\ {{\mathbf{r}-\mathbf{r}_p}\over {|\mathbf{r}-\mathbf{r}_p|^3}}\ dV_E - \mathcal{G} m_S \Bigl({{\mathbf{r}-\mathbf{r}_S}\over {|\mathbf{r}-\mathbf{r}_S|^3}}+{\mathbf{r}_S\over {|\mathbf{r}_S|^3}}\Bigr)\nonumber\\
&-& \mathcal{G} m_M\Bigl({{\mathbf{r}-\mathbf{r}_M}\over {|\mathbf{r}-\mathbf{r}_M|^3}}+{\mathbf{r}_M\over {|\mathbf{r}_M|^3}}\Bigr)+\mathbf{F}_{SRP}\ .
\eeqa
In \equ{eq1} we denote by $\mathcal{G}$ the gravitational constant, $\rho(\mathbf{r}_p)$ is the density at the point $\mathbf{r}_p$, $V_E$ is the volume of the Earth,
$m_S$, $m_M$ are the masses of the Sun and Moon, respectively, $\mathbf{r}_S$, $\mathbf{r}_M$ denote the position vectors
of the Sun and Moon with respect to the center of the Earth. Finally, $\mathbf{F}_{SRP}$ denotes
the effect of the solar radiation pressure (in short, SRP), which can be written as follows:
$$
\mathbf{F}_{SRP}=C_r\ P_r\ a_S^2\ \Bigl({A\over m}\Bigr)\ {{\mathbf{r}-\mathbf{r}_S}\over {|\mathbf{r}-\mathbf{r}_S|^3}}\ ,
$$
where $C_r$ is the reflectivity coefficient, depending on the optical properties of the surface of the space debris,
$P_r$ is the radiation pressure for an object located\footnote{$AU$ stands for Astronomical Unit;
$1\, AU$ corresponds to the average Earth-Sun distance, approximately equal to $1.496\cdot 10^8\ km$.}
at $a_S=1\, AU$, ${A\over m}$ is the area--to--mass ratio with $A$ the cross--section of
the space debris and $m$ its mass.

\vskip.1in

Let us denote by $\nabla_F$ and $\nabla_I$ the gradients in the synodic and quasi--inertial reference frames, respectively:
$$
\nabla_F\equiv {{\partial}\over {\partial X}}\mathbf{f}_1+{{\partial}\over {\partial Y}}\mathbf{f}_2+{{\partial}\over {\partial Z}}\mathbf{f}_3\ ,\qquad
\nabla_I\equiv {{\partial}\over {\partial x}}\mathbf{e}_1+{{\partial}\over {\partial y}}\mathbf{e}_2+{{\partial}\over {\partial z}}\mathbf{e}_3\ .
$$
The equations of motion \equ{eq1} can then be expressed as
\beqa{eq2}
\ddot{\mathbf{r}}=&-&R_3(-\theta)\ \nabla_F V(\mathbf{r})+ \mathcal{G} m_S\ \nabla_I\Bigl({1\over {|\mathbf{r}-\mathbf{r}_S|}}+{{\mathbf{r}\cdot \mathbf{r}_S}\over {|\mathbf{r}_S|^3}}\Bigr)\nonumber\\
&+&  \mathcal{G} m_M\ \nabla_I\Bigl({1\over {|\mathbf{r}-\mathbf{r}_M|}}+{{\mathbf{r}\cdot \mathbf{r}_M}\over {|\mathbf{r}_M|^3}}\Bigr)+
C_r\ P_r\ a_S^2\ {A\over m}\ {{\mathbf{r}-\mathbf{r}_S}\over {|\mathbf{r}-\mathbf{r}_S|^3}}\ ,
\eeqa
where the potential energy due to the Earth's attraction is given by
\beq{Vearth}
V(\mathbf{r})= -\mathcal{G}  \int_{V_E}{{\rho(\mathbf{r}_p)}\over {|\mathbf{r}-\mathbf{r}_p|}}\ dV_E\ .
\eeq

\subsection{Expression of the Earth's potential in spherical harmonics}\label{sec:VV}
In this section we express the potential energy $V$ in \equ{Vearth} in spherical harmonics.
To this end, in the synodic reference frame we can write
\beqano
X&=&r\cos\phi\cos\lambda\nonumber\\
Y&=&r\cos\phi\sin\lambda\nonumber\\
Z&=&r\sin\phi\ ,
\eeqano
where $(r,\lambda,\phi)$ are spherical coordinates with the longitude $0\leq\lambda< 2\pi$ and
the latitude $-{\pi\over 2}\leq \phi < {\pi\over 2}$. Let $R_E$ denote the Earth's equatorial radius.
The series expansion of the gravity geopotential in terms of the spherical harmonics is given by
\beq{potential}
V(r,\phi,\lambda)=-{{\mathcal{G} M_E}\over r}\ \sum_{n=0}^\infty \Bigl({R_E\over r}\Bigr)^n\ \sum_{m=0}^n P_{nm}(\sin\phi)\ (C_{nm}\cos m\lambda+
S_{nm}\sin m\lambda)\ .
\eeq
In \equ{potential} the quantities $P_{nm}$ are defined in terms of the Legendre polynomials $P_n(x)$ as
$$
P_{nm}(x)\equiv (1-x^2)^{m\over 2}\ {{d^m}\over {dx^m}}\{P_n(x)\}\ ,
$$
where $P_n(x)\equiv {1\over {2^n n!}}\ {{d^n}\over {dx^n}}\{(x^2-1)^n\}$.
The constants $C_{nm}$, $S_{nm}$ are defined as follows
\beqano
C_{nm}&\equiv&{{2-\delta_{0m}}\over M_E}\ {{(n-m)!}\over {(n+m)!}}\ \int_{V_E} \Bigl({r_p\over R_E}\Bigr)^n\ P_{nm}(\sin\phi_p)\
\cos(m\lambda_p)\rho(\mathbf{r}_p)\ dV_E\nonumber\\
S_{nm}&\equiv&{{2-\delta_{0m}}\over M_E}\ {{(n-m)!}\over {(n+m)!}}\ \int_{V_E} \Bigl({r_p\over R_E}\Bigr)^n\ P_{nm}(\sin\phi_p)\
\sin(m\lambda_p)\rho(\mathbf{r}_p)\ dV_E\ ,
\eeqano
where $M_E$ is the mass of the Earth, $(r_p,\lambda_p,\phi_p)$ denote the spherical coordinates associated to a point $P$ inside the Earth and,
again, $\mathbf{r}_p$ is its radius vector ($\delta_{jm}$ is the Kronecker symbol).

In the reference frame centered on the Earth's center of mass, one has that $C_{10}=C_{11}=S_{11}=0$;
moreover, following \cite{EGM2008} (see also Table~\ref{table:CS}), the values of  $C_{21}$ and $S_{21}$ are very small, so that we may neglect the contribution of these harmonics in the Cartesian equations. The expansion of the Earth's gravity potential up to $n=m=2$  is given by
\beqano
V(r,\phi,\lambda)
&\simeq&-{{\mathcal{G} M_E}\over r}-{{\mathcal{G} M_E}\over r}\ \Bigl({R_E\over r}\Bigr)^2\ \Big[ {1\over 2}(3\sin^2\phi-1)C_{20} \nonumber\\
& &+
3C_{22} \cos^2\phi\ \cos 2\lambda
+3S_{22} \cos^2\phi\ \sin 2\lambda\Big]\ .
\eeqano
The coefficients $C_{20}$ and $C_{22}$ can be written in the form:
$$
C_{20}={{A+B-2C}\over {2M_E R_E^2}}\ ,\qquad C_{22}={{B-A}\over {4M_E R_E^2}}\ ,
$$
where $A<B<C$ denote the principal moments of inertia.

In the synodic frame the potential can be written as
$$
V(X,Y,Z)=-{{\mathcal{G}M_E}\over r}-{{\mathcal{G}M_E}\over r}\ \Bigl({R_E\over r}\Bigr)^2\ \Big[C_{20} \Bigl({{3Z^2}\over {2r^2}}-{1\over 2}\Bigr)
+3C_{22} {{X^2-Y^2}\over r^2}+6S_{22}{{XY}\over r^2}\Big]\ .
$$
By using \eqref{catalin1}, we compute the first term of the right hand side of \equ{eq2} as
$$
 R_3(-\theta)\nabla_F V(r)= \Bigl({{\partial V}\over {\partial X}}\cos\theta-{{\partial V}\over {\partial Y}}\sin\theta \Bigr)\mathbf{e}_1+\Bigl({{\partial V}\over {\partial X}}\sin\theta+{{\partial V}\over {\partial Y}}\cos\theta \Bigr) \mathbf{e}_2+{{\partial V}\over {\partial Z}} \mathbf{e}_3\ .
$$
Introducing the notation
$$
C_S^-\equiv C_{22}\cos2\theta-S_{22}\sin2\theta\ ,\qquad C_S^+\equiv C_{22}\sin 2\theta+S_{22}\cos 2\theta\ ,
$$
we obtain the following equations of motion, with harmonics up to degree and order two:
\beqano
\ddot x&=&-{{\mathcal{G}M_Ex}\over r^3}+{{\mathcal{G}M_ER_E^2}\over r^5}\{C_{20}({3\over 2}x-{{15}\over 2}{{xz^2}\over r^2})+6 C_S^- x+6 C_S^+ y+
{{15 x}\over r^2}[C_S^-(y^2-x^2)-2xy C_S^+]\}\nonumber\\
&-&\mathcal{G}m_S\Bigl({{x-x_S}\over {|\mathbf{r}-\mathbf{r}_S|^3}}+{x_S\over r_S^3}\Bigr)-\mathcal{G}m_M \Bigl({{x-x_M}\over {|\mathbf{r}-\mathbf{r}_M|^3}}+{x_M\over r_M^3}\Bigr)+C_r\ P_r\ a_S^2\ \Bigl({A\over m}\Bigr)\ {{x-x_S}\over {|\mathbf{r}-\mathbf{r}_S|^3}}\nonumber\\
\ddot y&=&-{{\mathcal{G}M_Ey}\over r^3}+{{\mathcal{G}M_ER_E^2}\over r^5}\{C_{20}({3\over 2}y-{{15}\over 2}{{yz^2}\over r^2})+6 C_S^+ x-6 C_S^- y+
{{15 y}\over r^2}[C_S^-(y^2-x^2)-2xy C_S^+]\}\nonumber\\
&-&\mathcal{G}m_S\Bigl({{y-y_S}\over {|\mathbf{r}-\mathbf{r}_S|^3}}+{y_S\over r_S^3}\Bigr)-\mathcal{G}m_M \Bigl({{y-y_M}\over {|\mathbf{r}-\mathbf{r}_M|^3}}+{y_M\over r_M^3}\Bigr)+C_r\ P_r\ a_S^2\ \Bigl({A\over m}\Bigr)\ {{y-y_S}\over {|\mathbf{r}-\mathbf{r}_S|^3}}\nonumber\\
\ddot z&=&-{{\mathcal{G}M_Ez}\over r^3}+{{\mathcal{G}M_ER_E^2}\over r^5}\{C_{20}({9\over 2}z-{{15}\over 2}{{z^3}\over r^2})
+{{15 z}\over r^2}[C_S^-(y^2-x^2)-2xy C_S^+]\}\nonumber\\
&-&\mathcal{G}m_S\Bigl({{z-z_S}\over {|\mathbf{r}-\mathbf{r}_S|^3}}+{z_S\over r_S^3}\Bigr)-\mathcal{G}m_M \Bigl({{z-z_M}\over {|\mathbf{r}-\mathbf{r}_M|^3}}+{z_M\over r_M^3}\Bigr)+C_r\ P_r\ a_S^2\ \Bigl({A\over m}\Bigr)\ {{z-z_S}\over {|\mathbf{r}-\mathbf{r}_S|^3}}\ .
\eeqano

In a similar way, we may derive the contribution of the harmonics of any degree and order. In fact, for a specific problem,
we consider those Earth's gravity harmonics that are relevant for the given situation. For instance, in the study of the tesseral (or gravitational) resonances 1:1 and 2:1, we found that it is enough to expand the Earth's potential up to degree and order $n=m=3$
(\cite{CGmajor}). For different resonances, other harmonics might be relevant.

A list of some coefficients $C_{nm}$, $S_{nm}$ up to degree and order 4, as well as of the coefficients $J_{nm}$ computed through the formulae $J_{nm}=\sqrt{C_{nm}^2+S_{nm}^2}$ for $m>0$ and $J_{n0}=J_n=-C_{n0}$, is given in Table~\ref{table:CS} below,
which is derived from the EGM2008 model (\cite{EGM2008}, see also \cite{chao}).

\section{Delaunay variables}\label{sec:Delaunay}

The purpose of this section is to give a Hamiltonian formulation of the equations
of motion using a suitable set of action-angle variables and
taking into account the geopotential as well as the lunisolar
perturbations. In this Section we focus on objects with small area--to--mass ratio;
therefore, we disregard the influence of the solar radiation pressure, whose
investigation will be postponed to Section~\ref{sec:Am}.

 We introduce the
Delaunay variables $(L,G,H,M,\omega,\Omega)$, which can be defined in terms of
the orbital elements $\Upsilon=(a,e,i,M,\omega,\Omega)$ by means of the relations
\begin{equation}\label{Delaunay_variables}
L=\sqrt{\mu_E a}\ , \qquad G=L \sqrt{1-e^2}\ , \qquad H=G \cos i\ ,
\end{equation}
where $\mu_E=\mathcal{G} M_E$, $a$ is the semimajor axis, $e$ the eccentricity, $i$ the inclination, $M$ the mean anomaly, $\omega$ the argument of perigee, and $\Omega$ the longitude of the
ascending node. The orbital elements of the small body are referred to the celestial equator. The corresponding Hamiltonian can be written as
\beq{ham}
\mathcal{H}=-{\mu^2_E\over {2L^2}}+\mathcal{H}_{Earth}(\Upsilon,\theta)
-\mathcal{R}_{Sun}(\Upsilon,\Upsilon_S)-\mathcal{R}_{Moon}(\Upsilon, \Upsilon_M)\ ,
\eeq
where $\theta$ denotes the sidereal time, we denote by
$\Upsilon_S=(a_S,e_S,i_S,M_S,\omega_S,\Omega_S)$, $\Upsilon_M=(a_M,e_M,i_M,M_M,\omega_M,\Omega_M)$
the orbital elements of Sun and Moon, while $\mathcal{H}_{Earth}$, $\mathcal{R}_{Sun}$, $\mathcal{R}_{Moon}$ describe the perturbations due to the Earth, Sun and Moon, respectively.

The perturbations due to Sun and Moon will be made explicit in Sections~\ref{sec:RSun} and \ref{sec:RMoon}.
We remark that the orbital elements of Sun and Moon are known functions of time. In fact, the variation of the Sun's orbital elements
with respect to the celestial equator is well approximated by linear functions of time.
Concerning the Moon, its elements referred to the ecliptic (and not to the celestial equator) vary linearly in time (see \cite{gG74, HughesI, mL89}). Therefore, in modeling the lunisolar perturbations, it is important to use the equatorial elements of the Sun and the ecliptic elements of the Moon.

With this setting, the Hamiltonian $\mathcal{H}$ in \equ{ham} has
three degrees of freedom and an explicit time dependence. The associated canonical equations are:
\beqano
\dot{M}&=&\frac{\partial \mathcal{H}}{\partial L} \ , \qquad \dot{L}=-\frac{\partial \mathcal{H}}{\partial M} \nonumber\\
\dot{\omega}&=&\frac{\partial \mathcal{H}}{\partial G} \ , \qquad \dot{G}=-\frac{\partial \mathcal{H}}{\partial \omega}\nonumber\\
\dot{\Omega}&=&\frac{\partial \mathcal{H}}{\partial H} \ , \qquad \dot{H}=-\frac{\partial \mathcal{H}}{\partial \Omega}\ .
\eeqano

\subsection{The perturbation due to the Earth}\label{sec:REarth}
In the geocentric (quasi--inertial) frame, the Hamiltonian part $\mathcal{H}_{Earth}$ can be written as (see \cite{wK66})
\beq{RE}
\mathcal{H}_{Earth}=- {{\mu_E}\over a}\ \sum_{n=2}^\infty \sum_{m=0}^n \Bigl({R_E\over a}\Bigr)^n\ \sum_{p=0}^n F_{nmp}(i)\
\sum_{q=-\infty}^\infty G_{npq}(e)\ S_{nmpq}(M,\omega,\Omega,\theta)\ .
\eeq

The terms $F_{nmp}$ appearing in \equ{RE} are called the Kaula's inclination functions, and are defined by (see, e.g., \cite{wK62})
\beqa{Ffun}
F_{nmp}(i)&=&\sum_w {{(2n-2w)!}\over {w!(n-w)!(n-m-2w)!2^{2n-2w}}} \sin^{n-m-2w}i\ \sum_{s=0}^m\left(\begin{array}{c}
  m \\
  s \\
 \end{array}\right)
 \cos^si\nonumber\\
 &&\times \sum_c \left(\begin{array}{c}
  n-m-2w+s \\
  c \\
 \end{array}\right)
\left(\begin{array}{c}
  m-s \\
  p-w-c \\
 \end{array}\right)
 (-1)^{c-k}\ ,
\eeqa
where $k=[{{n-m}\over 2}]$, $[\cdot]$ denotes the integer part, the index $w$ runs between zero and the minimum between $p$ and $k$,
$c$ is taken over all values such that the binomial coefficients are not zero.
The functions $G_{npq}$, called eccentricity functions, are given by
\beq{Gfun}
G_{npq}(e)=(-1)^{|q|}(1+\beta^2)^n\beta^{|q|}\ \sum_{k=0}^\infty P_{npqk}Q_{npqk}\beta^{2k}\ ,
\eeq
where
$$
\beta={e\over {1+\sqrt{1-e^2}}}\ ,
$$
while
$$
P_{npqk}=\sum_{r=0}^h \left(\begin{array}{c}
  2p'-2n \\
  h-r \\
 \end{array}\right)
{{(-1)^r}\over r!}
\Bigl({{(n-2p'+q')e}\over {2\beta}}\Bigr)^r
$$
with $h=k+q'$ when $q'>0$ and $h=k$ when $q'<0$, and
$$
Q_{npqk}=\sum_{r=0}^h \left(\begin{array}{c}
  -2p' \\
  h-r \\
 \end{array}\right)
{1\over r!}
\Bigl({{(n-2p'+q')e}\over {2\beta}}\Bigr)^r\ ,
$$
where $h=k$ when $q'>0$ and $h=k-q'$ when $q'<0$,
$p'=p$ and $q'=q$ when $p\leq n/2$, $p'=n-p$ and $q'=-q$ when $p> n/2$.

In \equ{RE}, the quantities $S_{nmpq}$ are defined by
$$
S_{nmpq}=\left[%
\begin{array}{c}
  C_{nm} \\
  -S_{nm} \\
 \end{array}%
\right]_{n-m \ odd}^{n-m \ even} \cos \Psi_{nmpq}+ \left[%
\begin{array}{c}
  S_{nm} \\
  C_{nm} \\
 \end{array}%
\right]_{n-m \ odd}^{n-m \ even} \sin \Psi_{nmpq}\ ,
$$
where
$$\Psi_{nmpq}=(n-2p) \omega+(n-2p+q)M+m(\Omega-\theta)\ .$$

\begin{remark}
If we introduce the quantities $J_{nm}$ and $\lambda_{nm}$ defined through the relations
$$
J_{nm} = \sqrt{C_{nm}^2+S_{nm}^2}   \quad \textrm{if} \ m\neq 0\ , \qquad    J_{n0} \equiv J_n= -C_{n0}    \ ,$$
 $$C_{nm}=-J_{nm} \cos(m \lambda_{nm}) \ , \quad S_{nm}=-J_{nm} \sin(m \lambda_{nm}) \ ,
$$
then we can write $S_{nmpq}$ in the form
$$ S_{nmpq}=\left\{%
\begin{array}{cc}
  -J_{nm}  \cos \widetilde{\Psi}_{nmpq} \ , & \textrm{if} \ n-m\  \textrm{is even}\,, \\
  -J_{nm}  \sin \widetilde{\Psi}_{nmpq} \ , & \textrm{if} \ n-m\ \textrm{is odd} \,,\\
 \end{array}%
\right. $$
where
$$\widetilde{\Psi}_{nmpq}=(n-2p) \omega+(n-2p+q)M+m(\Omega-\theta)-m \lambda_{nm}\ .$$
\end{remark}

The values of $\lambda_{nm}$, up to degree and order 4, are given in Table~\ref{table:CS}.

\vskip.2in

\begin{table}[h]
\begin{tabular}{|c|c|c|c|c|c|}
  \hline
  $n$ & $m$ & $C_{nm}$ & $S_{nm}$ & $J_{nm}$ & $\lambda_{nm}$ \\
  \hline
  2 & 0 & -1082.6261& 0& 1082.6261& 0 \\
  2 & 1 & -0.000267& 0.0017873& 0.001807& $-81_{\cdot}^{\circ}5116$ \\
2 & 2 & 1.57462& -0.90387& 1.81559& $75_{\cdot}^{\circ}0715$ \\
 3 & 0 & 2.53241& 0& -2.53241& 0 \\
3 & 1 & 2.19315& 0.268087& 2.20947& $186_{\cdot}^{\circ}9692$ \\
3 & 2 & 0.30904& -0.211431& 0.37445& $72_{\cdot}^{\circ}8111$ \\
3 & 3 & 0.100583& 0.197222& 0.22139& $80_{\cdot}^{\circ}9928$ \\
4 & 0 & 1.6199& 0& -1.619331& 0 \\
 4 & 1 & -0.50864& -0.449265& 0.67864& $41_{\cdot}^{\circ}4529$ \\
4 & 2 & 0.078374& 0.148135& 0.16759& $121_{\cdot}^{\circ}0589$ \\
 4 & 3 & 0.059215& -0.012009& 0.060421& $56_{\cdot}^{\circ}1784$ \\
 4 & 4 & -0.003983& 0.006525& 0.007644& $-14_{\cdot}^{\circ}6491$ \\
  \hline
 \end{tabular}
 \vskip.1in
 \caption{A list of the coefficients $C_{nm}$, $S_{nm}$, $J_{nm}$ (in units of $10^{-6}$) and the quantities $\lambda_{nm}$; the values are computed from \cite{EGM2008}.}\label{table:CS}
\end{table}

\vskip.2in

\begin{remark}
As it is common in the literature, the terms of the expression \eqref{RE} are called {\sl tesseral} when $n \neq m$ and $m \neq 0$, they are named {\sl sectorial} when $n=m\neq 0$ and they are called {\sl zonal} when $n\neq 0$ and $m=0$.
\end{remark}

We provide now the analytical expansion of the solar and lunar disturbing functions appearing in \equ{ham}.

\subsection{The solar disturbing function $\mathcal{R}_{Sun}$}\label{sec:RSun}
When writing the gravitational potential due to the Sun, we can express the solar and satellite elements $\Upsilon_S=(a_S,e_S,i_S,M_S,\omega_S,\Omega_S)$ and $\Upsilon=(a,e,i,M,\omega,\Omega)$
with respect to the equator (see \cite{wK62}). We start by assuming that the Sun
moves on a Keplerian orbit
with semimajor axis $a_S=1\, AU$,
eccentricity $e_S=0.0167$, inclination $i_S=23^{\circ} 26' 21.406''$, argument of perigee $\omega_S=282.94^{\circ}$,
longitude of the ascending node $\Omega_S=0^{\circ}$. The mean anomaly changes as $\dot M_S\simeq 1^{\circ}/day$.

According to \cite{wK62}, the expansion of the gravitational solar potential is given by
\beqa{Rsun}
\mathcal{R}_{Sun}&=&\mathcal{G} m_S\sum_{l=2}^{\infty}\sum_{m=0}^l \sum_{p=0}^l \sum_{h=0}^l \sum_{q=-\infty}^\infty \sum_{j=-\infty}^\infty {a^l\over a_S^{l+1}}
\ \epsilon_m\, {{(l-m)!}\over {(l+m)!}}\nonumber\\
&&\times\ \F_{lmph}(i,i_S) \mathcal{H}_{lpq}(e)\, \mathcal{G}_{lhj}(e_S)\ \cos(\varphi_{lmphqj})\ ,
\eeqa
where
\beqano
\F_{lmph}(i,i_S)&\equiv&F_{lmp}(i)\ F_{lmh}(i_S)\ ,\nonumber\\
\varphi_{lmphqj}&\equiv& (l-2p)\omega+(l-2p+q)M-(l-2h)\omega_S-(l-2h +j)M_S+m(\Omega-\Omega_S)
\eeqano
with $m_S$ denoting the mass of the Sun; the quantities $\epsilon_m$ are defined as
$$
    \epsilon_m = \left\{ \begin{array}{cl} 1 & \text{if } m = 0\ , \\ 2 & \text{if } m \in\integer\backslash\{0\}\ , \end{array} \right.
$$
the functions $ \mathcal{H}_{lpq}(e)$ and $\mathcal{G}_{lhj}(e_S)$ are the Hansen coefficients $X_{l-2p+q}^{l,l-2p}(e)$, $X_{l-2h+j}^{-(l+1),l-2h}(e_S)$ (see \cite{giacaglia}), while the functions $F_{lmp}(i)$ and $F_{lmh}(i_S)$ are given in \equ{Ffun}.

We remark that the same expansion holds for the Moon (provided the solar elements
are replaced by the lunar ones), when the Moon's orbital elements are referred to the equator.

\subsection{The lunar disturbing function $\mathcal{R}_{Moon}$}\label{sec:RMoon}
Following \cite{gG74, HughesI, mL89}, the effect of the Moon on the satellite's
orbits is conveniently described when the orbital elements of the satellite are given with
respect to the equatorial plane and those of the
Moon with respect to the ecliptic plane.

Taking into account that the main
perturbing effect on the Moon is caused by the Sun, it turns out that the motion of the elements of the Moon
when referred to the celestial equator is nonlinear; in particular, the
changes of the argument of perigee and the longitude of the ascending node
are nonlinear. Indeed, the latter varies between $-13^{\circ}$ and
$+13^{\circ}$ with a period of 18.6 years. To bypass this problem, one
can express the lunar elements with respect to the ecliptic plane. In such case,
the inclination $i_M$ becomes nearly constant (in analogy to $a_M$ and
$e_M$), while the changes of the argument of perihelion
$\omega_M$ and that of the longitude of the ascending node $\Omega_M$ become
approximately linear (see for example \cite{Simon1994}) with rates of change, respectively, equal to
$\dot\omega_M\simeq 0.164^{\circ}/day$, $\dot\Omega_M\simeq
-0.053^{\circ}/day$, while the mean anomaly changes as $\dot M_M\simeq 13.06^{\circ}/day$. As a consequence,
the quantity $\omega_M+\Omega_M$ has a period of 8.85 years, while $\Omega_M$ varies with a period of 18.6 years.

To write the potential due to the Moon, let us start with the assumption that
the trajectory of the Moon is a Keplerian ellipse with semimajor axis $a_M=384\,748\, km$, eccentricity
$e_M=0.0549$ and inclination $i_M=5^{\circ}15'$.
The expansion of the lunar disturbing function is given, e.g., in \cite{CGPRnote}
(see also \cite{mL89}), as follows:
\beqa{Rgood}
    \mathcal{R}_{Moon}
    \nonumber
    & = &\mathcal{G} m_M\sum\limits_{l \geq 2} \sum\limits_{m = 0}^l \sum\limits_{p = 0}^l \sum\limits_{s = 0}^l
        \sum\limits_{q = 0}^l \sum\limits_{j = -\infty}^{+\infty} \sum\limits_{r = -\infty}^{+\infty}
        (-1)^{m+s}\ (-1)^{k_1} \frac{ \epsilon_m \epsilon_s}{2 a_M} \frac{(l - s)!}{(l + m)!}
        \left( \frac{a}{a_M} \right)^l \\
    \nonumber
    &\times& F_{lmp} (i) F_{lsq} (i_M)
        \mathcal{H}_{lpj} (e) \mathcal{G}_{lqr} (e_M) \\
    \label{eq:lane}
    &\times& \left\{ (-1)^{k_2} U_l^{m, -s}
        \cos \left( \bar{\theta}_{lmpj} + \bar{\theta}_{lsqr}^\prime - y_s \pi \right)
        + (-1)^{k_3} U_l^{m, s} \cos \left( \bar{\theta}_{lmpj} - \bar{\theta}_{lsqr}^\prime - y_s \pi \right)  \right\}\ ,\nonumber\\
\eeqa
where $y_s=0$ for $s$ even and $y_s=1/2$ when $s$ is odd, $k_1=[m/2]$, $k_2 = t (m + s - 1) + 1$, $k_3 = t (m + s)$ with $t=(l-1)$ mod 2,
the quantities $\bar{\theta}_{lmpj}$, $\bar{\theta}_{lsqr}^\prime$ are given by
\beqano
    \bar{\theta}_{lmpj} & = & (l - 2p) \omega + (l - 2p + j) M + m \Omega\ , \nonumber\\
    \bar{\theta}_{lsqr}^\prime & = & (l - 2q) \omega_M + (l - 2q + r) M_M + s (\Omega_M - \pi/2)\ ,
\eeqano
while the functions $U_l^{m,s}$ are defined as (compare with \cite{CGPRnote})
\beqano
U_l^{m,s}&=&\sum_{r =\max(0,-(m+s))}^{\min(l-s,l-m)} (-1)^{l-m-r}
\left(\begin{array}{c}
  l+m \\
  m+s+r \\
\end{array}\right)\
\left(\begin{array}{c}
  l-m \\
  r \\
\end{array}\right)\
\cos^{m+s+2r}({\varepsilon\over 2})\sin^{-m-s+2(l-r)}({\varepsilon\over 2})\ ,\nonumber\\
\eeqano
where $z=\cos^2 ({\varepsilon\over 2})$ and  $\varepsilon$ denotes the obliquity of the ecliptic, which is equal to $\varepsilon=23^{\circ}26'21.45''$

The functions $F_{lmp}(i)$ and $F_{lmh}(i_M)$ have been introduced in \equ{Ffun}, while $ \mathcal{H}_{lpj}(e)$ and $\mathcal{G}_{lqr}(e_S)$ are the Hansen coefficients $X_{l-2p+j}^{l,l-2p}(e)$, $X_{l-2q+r}^{-(l+1),l-2q}(e_M)$.

\subsection{Effects of $J_2$}\label{J2_effects}

From the infinite number of harmonic terms of the series expansions \eqref{RE}, \eqref{Rsun} and \eqref{Rgood}, only a few are really important for a specific
case study. When dealing with resonant motions, these terms are discriminated by analysing the cosine arguments and averaging over the fast angles, as well as by determining the dominant  harmonic terms.

To have a rough estimate of the rate of variation of the Delaunay angles, which makes possible the analysis of cosine arguments of the Fourier expansions \eqref{RE}, \eqref{Rsun} and \eqref{Rgood}, it is important to recall which are the principal secular effects of the $J_2$ zonal harmonic
(compare with \cite{wK66, KH1958}).

In MEO, for small area--to--mass ratio objects,  the perturbations due to the Earth's oblateness $J_2$ are at least one order of magnitude larger than the other effects. As a consequence, the secular effects on space debris orbits may be estimated with a good enough accuracy by a
simplified model described just by a single term (of the  order of $J_2$) in the series expansion \eqref{RE}, namely the term
corresponding to $n=2$, $m=0$, $p=1$, $q=0$.

We stress that $J_2$ plays a very important role not only in estimating the secular effects upon the space debris orbits, but also
for being responsible of both, the occurrence of the phenomenon of overlapping of tesseral resonances and the occurrence of lunisolar secular resonances.
As it will be shown in Section~\ref{sec:chaos},  $J_2$ has a very clear role in the chaotic variation of the orbital elements, in particular the semimajor
axis and the eccentricity.

In the light of the previous discussion, we find important to recall the solution associated to the Hamiltonian obtained by summing the Keplerian part $-\mu_E^2/(2L^2)$ and the
secular part (namely, the term of the series \eqref{RE} for which $n=2$, $m=0$, $p=1$, $q=0$). Since  $F_{201}(i)=0.75 \sin^2 i -0.5$ and $G_{210}(e)=(1-e^2)^{-3/2}$ (see \eqref{Ffun} and \eqref{Gfun}), from  \eqref{Delaunay_variables} we obtain the Hamiltonian
$$
\H_{Kepler+J_2}=-\frac{\mu_E^2}{2 L^2}+{{R_E^2 J_2 \mu_E^4}\over {4}}\ {{1}\over {L^3G^3}}\ \Bigl(1-3{H^2\over G^2}\Bigr)\ .
$$
Since the angles in $\H_{Kepler+J_2}$ are ignorable, it follows that $L$, $G$ and $H$ are constant, while the Delaunay variables $M$, $\omega$ and $\Omega$ vary linearly in time with rates
\beqa{elements}
\dot{M}  &=&  \frac{\mu_E^2}{L^3}-\frac{3 R_E^2 J_2 \mu_E^4}{4} \frac{1}{L^4 G^3} \Bigl(1-3 \frac{H^2}{G^2}\Bigr)\,,\nonumber\\
\dot{\omega} &=&  \frac{3 R_E^2 J_2 \mu_E^4}{4} \frac{1}{L^3 G^4} \Bigl(-1+5 \frac{H^2}{G^2}\Bigr)\,,\nonumber\\
\dot{\Omega} &=& - \frac{3 R_E^2 J_2 \mu_E^4}{2} \frac{H}{L^3 G^5}\ .
\eeqa
By using the relations \eqref{Delaunay_variables} and the fact that $\mu_E^{1/2} R_E^{-3/2} \simeq 107.1 \, [day^{-1}]$, we can rewrite the
relations \equ{elements} in terms of the orbital elements as:
\begin{equation}
\begin{split} \label{MomeagaOmega_var}
& \dot{M}  \simeq 6135.7 \Bigl(\frac{R_E}{a}\Bigr)^{3/2}- 4.98 \Bigl(\frac{R_E}{a}\Bigr)^{7/2} (1-e^2)^{-3/2} (1-3 \cos^2 i) \ ^o/day\,,\\
& \dot{\omega} \simeq  4.98 \Bigl(\frac{R_E}{a}\Bigr)^{7/2} (1-e^2)^{-2} (5 \cos^2 i-1) \ ^o/day\,,\\
& \dot{\Omega} \simeq - 9.97 \Bigl(\frac{R_E}{a}\Bigr)^{7/2} (1-e^2)^{-2}  \cos i \ ^o/day\,. \\
\end{split}
\end{equation}
Therefore, we are led to summarize as follows the main effects of $J_2$: a slow change of the rate of the mean anomaly, a precession of the perigee and a secular regression of the orbital node.

\section{Resonances}\label{sec:resonances}

In \eqref{RE}, \eqref{Rsun} and \eqref{Rgood} we have expressed the perturbations due to the Earth, Sun and Moon as series involving an infinite number of combinations of the following angles: $\theta$, $M$, $M_b$, $\omega$, $\omega_b$, $\Omega$ and $\Omega_b$, with $b=S,M$. From all possible combinations, relevant for the long--term evolution of the orbital elements are only those which involve some specific cosine arguments. The degree of influence of the harmonic terms in the series expansions depends on the explored region and the scale of time for which the dynamics is studied.

We can classify all arguments by considering the frequencies (or periods) associated with the cosine arguments in the expansions \eqref{RE}, \eqref{Rsun} and \eqref{Rgood}. Thus,  since the frequency of $\theta$ is equal to a rotation per sidereal day, in view of \eqref{MomeagaOmega_var} and
Sections~\ref{sec:RSun} and \ref{sec:RMoon}, it follows that the angles $M$ and $\theta$ are rapidly varying quantities, $M_M$ and $M_S$ change with constant moderate rates, $\Omega_S$ and $\omega_S$ may be considered constant, while $\omega$, $\omega_M$, $\Omega$ and $\Omega_M$ undergo slow variations.

Comparing the space debris problem with some classical models in Celestial Mechanics, for instance the Sun--Jupiter--asteroid
three-body problem, we notice that there is an important difference between the time scales of variation of the angles involved
in the expansions of the corresponding disturbing functions.
For the Sun--Jupiter--asteroid system (see, e.g., \cite{cMsD99}), one has two types of angles:
fast angles (the mean anomalies of the perturbed and perturbing bodies) and slow angles (the argument of pericentre and the longitude of the ascending node for both, the asteroid and Jupiter).
The terms of the disturbing function for the three-body problem may be classified as: secular (independent of the fast angles), short periodic (involve fast angles) and resonant (imply commensurabilities between the fast angles).
In the context of the space debris problem, one has three categories of angles: fast angles ($M$ and $\theta$), semi--fast angles ($M_M$ and $M_S$) and slow angles ($\omega$, $\omega_M$, $\Omega$ and $\Omega_M$).
Although this classification of the angles is rather conventional, it gives a clear idea, at least
within\footnote[10]{In LEO, the rate of variation of the angles $\omega$ and $\Omega$ is comparable with those of
the angles $M_S$ and $M_M$ (see \eqref{MomeagaOmega_var}, where $R_E/a$ is close to one).} MEO and GEO, on the rate of variation of the angles appearing as cosine arguments of the expansions.

In the light of the above discussion, we may adopt a classification of the terms of the expansions \eqref{RE}, \eqref{Rsun} and \eqref{Rgood} as follows.
Arguments that do not depend on the sidereal time $\theta$ and the mean anomalies $M$, $M_S$, $M_M$ give rise to {\it secular terms}.
If the trigonometric arguments are independent of $\theta$ and $M$, but depend on either $M_S$ or $M_M$, then we speak about {\it semi--secular terms}. Finally, the arguments involving the fast angles give rise either
to {\it resonant terms}, when there is a commensurability between $\dot{\theta}$ and $\dot{M}$, or to {\it short periodic terms},
when $\dot{\theta}$ and $\dot{M}$ are not commensurable.

Resonances involving commensurabilities between the Earth's rotation period and
the orbital period of the space debris (\cite{BWM, CGmajor, CGminor, VDLC}) are called \sl tesseral (or gravitational) resonances\rm .
It is remarkable that tesseral resonances provoke variations of the semimajor axis on a time scale of the order of hundreds of days. Their effects have been studied in~\cite{BWM, CGmajor, CGminor, CGext, CGleo, VDLC}, within LEO, MEO and GEO, as well as outside the geostationary ring.

Beside tesseral resonances, another class of commensurabilities,
called \sl lunisolar resonances\rm , affects the motion of space debris (and artificial satellites). In principle, following \cite{HughesI},  there are  15 possible types of third-body (lunar and solar) resonances.
This classification accounts for all possible resonances involving a third--body perturber: \sl secular resonances, semi--secular resonances \rm and \sl mean motion resonances.\rm \,
However, within third--body resonances, only some specific resonances affect the dynamics.

Since in a given orbital elements' region the mean motion resonances occur when the ratio of the orbital periods of the perturbed and perturbing bodies
equals a rational number, such resonances do not occur in LEO, MEO or GEO, but rather at a much larger distance from the center of the Earth.  Therefore, the mean motion resonances are
less interesting for the space debris problem.

The same situation holds for the semi--secular resonances, at least in MEO and GEO. Indeed, since the semi--secular resonances involve the mean anomalies of the Moon and Sun, whose rates of variation are $\dot M_M\simeq 13^{\circ}/day$, $\dot M_S\simeq 1^{\circ}/day$, it turns out that these resonances mostly take place in the LEO region (see Section~\ref{sec:other}).
Since the semi--secular resonances
occur at specific altitudes, the width of such resonances is small, and since
the air drag provokes a decay of the orbits on relatively short time scales, one expects that these resonances  play a minor role in the long--term evolution of space debris.

On the contrary, the secular resonances are of seminal importance in understanding the global dynamics of space debris, in particular in the MEO region. Involving commensurabilities among the slow frequencies
of the orbital precession of a satellite and the perturbing body
(\cite{CGPRnote, DRADVR15, ElyHowell, GDGR, HughesI, aR15}), the secular resonances influence the evolution of the eccentricity
and inclination on time scales of the order of tens (or hundreds) of years. In addition, the phenomenon of overlapping
of secular resonances possibly contributes to the design of disposal orbits (\cite{RDAVRD}). Although in a few years the MEO region will be populated by four complete constellations of satellites, namely GPS, GLONASS, Galileo and BeiDou,
there are no internationally agreed mitigation guidelines as it happens within LEO and GEO (see~\cite{IADC}).
However, it is important to stress that in the neighborhood of the above mentioned navigation satellite systems in MEO,
the secular resonances give rise to an intricate dynamics, with a consequent difficulty in choosing the best disposal scenario.

\subsection{Tesseral resonances} \label{sec:tesseral_resonances}
Once we have listed the possible resonances that affect the motion of a space debris, we discuss now how we can construct a simplified model
that describes, with a good enough accuracy, the resonant dynamics. We focus first on tesseral resonances, which are formally defined
as follows.

\vskip.1in

\begin{definition}
A $j:\ell$ tesseral (or gravitational) resonance with $j$, $\ell\in\integer_+$ occurs whenever the orbital period of the debris
and the rotational period of the Earth are commensurable of order $j/\ell$. Expressed in terms of the orbital elements,
we have a $j:\ell$ gravitational resonance, whenever the following relation is satisfied:
\beq{resdef}
\ell\ \dot{M}-j\ \dot{\theta} = 0\ , \qquad j,\ell \in \integer_+\ .
\eeq
\end{definition}

We remark that in concrete astronomical cases, the expression \equ{resdef} is satisfied within a
definite accuracy and it cannot be satisfied exactly.

From Kepler's third law, a $j:\ell$ resonance corresponds to the semimajor axis
$a_{j:\ell}=(j/\ell)^{-2/3}\ a_{geo}$, where $a_{geo} = 42\,164.17\, km$
denotes the semimajor axis of the geosynchronous orbit. We report in
Table~\ref{table:J} an estimate of the position of different resonances as derived from
Kepler's third law. In fact, due to perturbations, in particular the influence of $J_2$, the exact position
of the resonances depends of the values of the orbital elements (see \cite{CGminor}).

\vskip.2in

\begin{table}[h]
\begin{tabular}{|c|c||c|c||c|c|}
  \hline
  $j:\ell$ & $a$ in km & $j:\ell$ & $a$ in km\\
  \hline
 1:1 & 42\,164.2 & 4:3 & 34\,805.8 \\
 2:1 & 26\,561.8 & 1:2 & 66\,931.4\\
 3:1 & 20\,270.4 & 1:3 & 87\,705.0\\
 3:2 & 32\,177.3 & 2:3 & 55\,250.7\\
 4:1 & 16\,732.9 &  & \\
 \hline
 \end{tabular}
 \vskip.1in
 \caption{Semimajor axis $a$ of some gravitational resonances of order $j:\ell$ with $j$, $\ell \in \mathbb{Z}_+$.}\label{table:J}
\end{table}

\vskip.2in

A key point in the study of resonant dynamics is the analysis of only those harmonic terms in the expansions \eqref{RE}, \eqref{Rsun} and \eqref{Rgood} that really count for a given $j:\ell$ tesseral resonance.
In this way, the computational effort is considerably decreased and the global dynamics is more clearly understood.
The procedure for the selection of the dominant terms is the following.

First, we identify the resonant terms. To this end, it is worth to underline that the expansions of the lunar and solar potentials do not contain resonant terms. Indeed, since \eqref{Rsun} and \eqref{Rgood} are
independent on the sidereal time $\theta$, the long--term variation of the semimajor axis is directly affected by the resonant part of the geopotential.
According to the averaging principle (see, e.g., \cite{cMsD99}),
the effects of the short periodic terms average out over a long time-scale.
Hence, such terms can be dropped from the expansions \eqref{RE}, \eqref{Rsun} and \eqref{Rgood}, and we can focus only on the most
relevant terms. We drop also the semi-secular terms from the expansions,  since their effects average out over a long enough interval of time, but also
because they do not affect directly the variation of the semimajor axis. Moreover, their magnitude is much smaller than that of the most influent secular terms, at least in MEO.

Then, we compute the secular part of the Hamiltonian. This part influences also the long--term behavior of the semimajor axis, though indirectly.  Being independent on the mean anomaly $M$, the secular part is not involved explicitly in the canonical equation describing the evolution of $L$ (or the semimajor axis).
However, the long--term evolution of the other orbital elements (equivalently, the other Delaunay variables) is directly affected by the secular Hamiltonian. In particular, as effect of $J_2$ (see Section~\ref{J2_effects}), the argument of perigee $\omega$ and the longitude of the ascending node $\Omega$ vary slowly over time and therefore they indirectly affect the evolution of the semimajor axis, since the canonical equation describing the evolution of the semimajor axis involves all orbital elements as parameters. In MEO, due to the variation of  $\omega$, described by \eqref{MomeagaOmega_var},  the
phenomena of splitting or overlapping of tesseral resonances occurs (see \cite{CGminor}). Such phenomena  lead to chaotic variations of the semimajor axis. The secular part due to the Moon and Sun has also an indirect role in the variation of semimajor axis, since it induces basically a long--term variation of the eccentricity and inclination. However, the time scale of variation is totally
different: the semimajor axis varies on a time scale of the order of hundreds of days,
while the evolution of the eccentricity
and inclination occurs on a much longer time scale, of the order of tens (or hundreds) of years. For regions located outside the geostationary ring, the influence of the Moon and Sun is larger than that of $J_2$, and  thus  $\omega$ and $\Omega$ vary  as effect of the secular part due to the Moon and Sun, rather than under the influence of $J_2$.

In \cite{CGmajor, CGminor}, by comparing the results obtained within the Hamiltonian framework with those provided by the Cartesian model described in Section~\ref{sec:model},
it was noticed that for the tesseral resonances located in MEO and GEO, one can rely on a simplified model that disregards the secular part corresponding to
Sun and Moon. On the contrary, for the tesseral resonances located outside the geostationary ring (see \cite{CGext}), one should consider the effects of
Sun and Moon. In the following, for completeness, we describe both the secular part due to the Earth's potential and that due to the Moon and Sun.

Finally, in order to reduce the number of harmonics to just those which really shape the dynamics, we perform an analysis of
the dominant terms in specific regions of the orbital element's space. This procedure is described in detail in \cite{CGmajor} and applied to
some tesseral resonance studied in~\cite{CGminor, CGext}.

In practical computations, the analysis of dominant terms allows us to reduce the discussion to a limited number of terms and to provide an indication of the optimal degree of expansion. More precisely, for a given $j:\ell$ resonance, we  truncate $\H_{Earth}$
up to an optimal degree $N$, namely we consider a finite
series expansion of the form
\beq{R1}
\H_{Earth}=\H^{sec}_{Earth}+\H_{Earth}^{res\, j:\ell}+\H_{Earth}^{nonres}\cong
\sum_{n=2}^N \sum_{m=0}^n \sum_{p=0}^n \sum_{q=-\infty}^{\infty} \mathcal{T}_{nmpq} \ ,
\eeq
where we split the Hamiltonian part $\H_{Earth}$  into the secular part $\H^{sec}_{Earth}$, the resonant contribution $\H_{Earth}^{res\, j:\ell}$,
and the non-resonant part $\H_{Earth}^{nonres}$. In formula \equ{R1} we have introduced the coefficients $\mathcal{T}_{nmpq}$
defined by
$$
\mathcal{T}_{nmpq}=-\frac{\mu_E R_E^n}{a^{n+1}} F_{nmp}(i)G_{npq}(e) S_{nmpq}(M, \omega, \Omega , \theta)\ .
$$
The secular part due to the Earth's potential is obtained by taking the terms corresponding to $m=0$ and $n-2p+q=0$. Although one may limit to
the $J_2$ approximation of the secular part, for $|J_2|\gg |J_n|$, $n>2$, in practical computations we consider the expansion of the secular part up to $n=4$.   Thus, up to the second order
in the eccentricity, we obtain the following expression:
\beqa{Rsec}
\H_{Earth}^{sec}&\cong&\frac{\mu_E R^2_E J_{2}}{a^3} \Bigl(\frac{3}{4} \sin^2 i -\frac{1}{2}\Bigr) (1-e^2)^{-3/2} \nonumber\\
&+&\frac{2\mu_E R^3_E J_{3}}{a^4} \Bigl(\frac{15}{16} \sin^3 i -\frac{3}{4} \sin i\Bigr) e (1-e^2)^{-5/2} \sin \omega \nonumber \\
&+&\frac{\mu_E R^4_E J_{4}}{a^5} \Bigl[\Bigl(-\frac{35}{32} \sin^4 i +\frac{15}{16} \sin^2 i\Bigr) \frac{3e^2}{2}(1-e^2)^{-7/2} \cos(2\omega) \nonumber \\
&+&
\Bigl(\frac{105}{64} \sin^4 i -\frac{15}{8} \sin^2 i+\frac{3}{8}\Bigr) (1+\frac{3e^2}{2})(1-e^2)^{-7/2} \Bigr]\ .
\eeqa

\vskip.1in

The resonant part corresponding to a resonance of order $j:\ell$ is obtained by retaining the terms $\mathcal{T}_{nmpq}$ for which
$j(n-2p+q)=\ell m$. Expansions for specific low-order resonances can be found, e.g., in \cite{CGmajor, CGminor, CGext}. As an example, we provide here the expansion up to second order in the eccentricity
and fourth power of $R_E$ of the 2:1 resonance:
\begin{equation}
\begin{split}\label{H21}
& \H_{Earth}^{res\, 2:1}\cong \frac{\mu_E R_E^2 J_{22}}{a^3} \Bigl\{-\frac{3}{8} e (1+\cos i)^2 \cos (M-2\theta +2\omega +2\Omega -2\lambda_{22})\\
& \qquad + \frac{9}{4} e \sin^2 i \cos (M-2\theta +2\Omega -2\lambda_{22})
\Bigr\}\\
& \ + \frac{\mu_E R_E^3 J_{32}}{a^4} \Bigl\{ \frac{15}{64} e^2 \sin i (1+\cos i)^2 \sin (M-2\theta +3 \omega +2\Omega -2\lambda_{32}) \\
& \qquad + \frac{15}{8} \Bigl(1+2e^2 \Bigr) \sin i (1-2\cos i-3 \cos^2 i)  \sin (M-2\theta + \omega +2\Omega -2\lambda_{32}) \\
&  \qquad - \frac{165}{64} e^2 \sin i (1+2\cos i-3 \cos^2 i) \sin (M-2\theta - \omega +2\Omega -2\lambda_{32}) \Bigr\}\\
 & \ + \frac{\mu_E R_E^4 J_{42}}{a^5} \Bigl\{ \frac{e}{2} \Bigl(\frac{105}{8} \sin^2 i \cos i (1+\cos i) -\frac{15}{8} (1+\cos i)^2 \Bigr) \cos(M-2\theta + 2\omega+ 2\Omega -2\lambda_{42})\\
& \qquad + \frac{5 e}{2} \Bigl(\frac{105}{16} \sin^2 i (1-3\cos^2 i) -\frac{15}{4} \sin^2 i \Bigr) \cos(M-2\theta + 2\Omega -2\lambda_{42}) \Bigr\}\\
& \ + \frac{\mu_E R_E^4 J_{44}}{a^5} \Bigl\{ \frac{105}{32} e^2 (1+\cos i)^4 \cos [2(M-2\theta + 2\omega +2\Omega -2\lambda_{44})] \\
& \qquad + \frac{105}{4} \Bigl(1+e^2 \Bigr) \sin^2 i (1+ \cos i)^2  \cos [2(M-2\theta + \omega +2\Omega -2\lambda_{44})]\\
& \qquad + \frac{1575}{8} e^2 \sin^4 i \cos [2(M-2\theta + 2\Omega -2\lambda_{44})]\Bigr\}\ .
 \end{split}
\end{equation}

In conclusion, a simplified model able to describe the main dynamical features of the $j:\ell$ resonance is provided by the Hamiltonian
$$
\H_{Kepler+Earth}^{j:\ell}=-\frac{\mu_E^2}{2L^2} + \H_{Earth}^{sec}+\H_{Earth}^{res \, j:\ell}\,,
$$
which is obtained from the general Hamiltonian \eqref{ham} by retaining the relevant secular and resonant harmonic terms from the expansion of the Earth's potential. This model was validated in \cite{CGmajor, CGminor, CGext} by performing  numerical integrations in Cartesian variables of the more compete model including the gravitational attraction of the Sun, Moon as well as solar radiation pressure (see Section~\ref{sec:model}).

As mentioned above, a more complete Hamiltonian, allowing to model also lunisolar secular resonances, is obtained by including the secular part of the expansions \eqref{Rsun} and \eqref{Rgood}.
We stress again that, as effect of tesseral resonances, the semimajor axis varies on a time scale of the order of hundreds of days,
while, due to secular resonances, the evolution of the eccentricity
and inclination occurs on a much longer time scale, of the order of tens (or hundreds) of years.
The  model including the secular contributions of Sun and Moon is described by the Hamiltonian
\begin{equation}\label{ham_tess_res_all}
\H^{j:\ell}=-\frac{\mu_E^2}{2L^2} + \H_{Earth}^{sec}+ \H_{Sun}^{sec}+ \H_{Moon}^{sec}+\H_{Earth}^{res \, j:\ell}\ ,
\end{equation}
where
\begin{equation} \label{Rb_versus_Hb}
\H_{Sun}^{sec}=-\overline{\mathcal{R}}_{Sun}, \qquad  \H_{Moon}^{sec}=-\overline{\mathcal{R}}_{Moon}\,,
\end{equation}
and $\overline{\mathcal{R}}_{Sun}$ and $\overline{\mathcal{R}}_{Moon}$ are obtained by averaging $\mathcal{R}_{Sun}$ and $\mathcal{R}_{Moon}$, respectively, over both the mean anomaly of the small body and the mean anomaly of the third--body perturber.

Considering the expansions \eqref{Rsun} and \eqref{Rgood} up to degree 2 in the ratio of semimajor axes and averaging over the mean anomalies $M$ and $M_b$, $b=S,M$,
one obtains (see also \cite{CGPbif}):
\begin{equation}
\begin{split}\label{Rsun_double_average}
& \overline{\R}_{Sun}=\G m_S\sum_{m=0}^2 \sum_{p=0}^2 {a^2\over a_S^3}\, \frac{1}{(1-e_S)^{3/2}}
\ \epsilon_m\, {{(2-m)!}\over {(2+m)!}}\ F_{2mp}(i)\ F_{2m1}(i_S)\\
&\qquad \times X_0^{2,\, 2-2p}(e)\,\ \cos\Bigl((2-2p)\omega+m(\Omega-\Omega_S)\Bigr)\ ,
\end{split}
\end{equation}
and
\begin{equation}
\begin{split}\label{Rmoon_double_average}
& \overline{\R}_{Moon}={1\over 2}\ \G m_M\sum_{m=0}^2 \sum_{s=0}^2 \sum_{p=0}^2  {a^2\over a_M^3}\,  \frac{1}{(1-e_M)^{3/2}}\,
(-1)^{[{m\over 2}]}\ \epsilon_m\, \epsilon_s\ {{(2-s)!}\over {(2+m)!}}\\
&\qquad \times F_{2mp}(i)\ F_{2s1}(i_M)\ X_0^{2,\, 2-2p}(e)\, \\
&\qquad  \times \Big\{ U_2^{m,-s} \cos\Bigl((2-2p)\omega+ m\Omega+s\Omega_M-s{\pi\over 2}-y_s\pi\Bigr)\\
&\qquad + U_2^{m,s} \cos\Bigl((2-2p)\omega+m\Omega-s\Omega_M+s{\pi\over 2}-y_s\pi \Bigr)\Big\}\ .
\end{split}
\end{equation}

\subsection{Secular resonances}\label{sec:inclination}

Provoked by the complex interaction between the Earth's oblateness and the gravitational
attraction of Sun and Moon, the secular resonances are of great importance in the long--term stability of satellites and space debris, in particular in the MEO region where
the satellite navigation systems are located.
Such resonances occur whenever there is a commensurability between the slow frequencies of orbital precession of the debris and the perturbing body (see \cite{HughesI}).

Although the variation of the orbital elements of an Earth's satellite under the gravitational attraction of
Sun and Moon is investigated since the mankind started the conquest of the space
(see \cite{cook1962, wK62, Kozai1959, Kozai66, musen1961, upton1959}), the effects of secular resonances are not yet completely understood, and they are still
largely studied both qualitatively and quantitatively (see
\cite{ADRRVDM2016, Breiter2001, CGfrontier, CGPbif, CGPRnote, chaogick, DRADVR15, deleflie2011,
GDGR,  RDGF2015, aR15, aR08, Sanchez15, zhuetal2015}). The growth in eccentricity, which may decrease the perigee altitude
even down to the LEO region, the overlapping of nearby secular resonances, the bifurcation of equilibria are only
some of the dynamical phenomena that can occur in the physical model, and that are definitely important
in astrodynamics, as well as from the mathematical point of view.

Following \cite{Breiter2001, CGPbif, HughesI, aR08} we introduce the following definition of a
lunar and solar \sl gravity secular resonance. \rm

\begin{definition} \label{def:secres}

A solar gravity secular resonance occurs whenever there exist $(k_1,k_2,k_3,k_4)\in\integer^4\backslash\{0\}$, such that
\beq{secressun}
k_1\dot\omega+k_2\dot\Omega+k_3\dot\omega_S+k_4\dot\Omega_S=0\ .
\eeq
We have a lunar gravity secular resonance whenever there exists a vector
$(k_1,k_2,k_3,k_4)\in\integer^4\backslash\{0\}$, such that
\beq{secresmoon}
k_1\dot\omega+k_2\dot\Omega+k_3\dot\omega_M+k_4\dot\Omega_M=0\ .
\eeq

\end{definition}

We stress that the above definition of secular resonance is as general as possible.
However, given the fact that in MEO the lunar and solar expansions may be reduced to the simplified forms \eqref{Rsun_double_average} and \eqref{Rmoon_double_average}, that is they are truncated to the second order in the ratio of semimajor axes and averaged over $M$ and $M_b$, with $b=S,M$, the Hamiltonian
turns out to be independent of $\omega_M$ and $\omega_S$.
Therefore, one has $k_3=0$ in  \eqref{secressun} and \eqref{secresmoon}. Moreover, since $\dot{\Omega}_S \simeq 0$,  the relations \eqref{secressun} and \eqref{secresmoon}
associated to \equ{Rsun_double_average} and \equ{Rmoon_double_average} can be written in the form:
\beq{secressunbis}
(2-2p)\dot\omega+m\dot\Omega=0\ , \qquad m,p=0,1,2\,,
\eeq
and
\beq{secresmoonbis}
(2-2p)\dot\omega+m\dot\Omega+s \dot\Omega_M=0\ , \qquad m,p=0,1,2, \quad s=-2,-1,0,1,2\,,
\eeq
respectively.

Provided the region of interest is outside the libration region of a $j: \ell$ tesseral resonance\footnote{The libration island associated to a tesseral resonance does not exceed 100 $km$ in width (compare with \cite{CGmajor, CGminor}).}, one may reduce the problem to the following two
degrees of freedom non-autonomous Hamiltonian:
\begin{equation}\label{ham_sec_res_all}
\H^{sec}= \H_{Earth}^{sec}+ \H_{Sun}^{sec}+ \H_{Moon}^{sec}\,,
\end{equation}
where $\H_{Earth}^{sec}$, $\H_{Sun}^{sec}$ and $\H_{Moon}^{sec}$ are defined by the relations \eqref{Rsec}, \eqref{Rb_versus_Hb}, \eqref{Rsun_double_average} and \eqref{Rmoon_double_average}. In fact, as it was noted in
\cite{CGfrontier, CGPbif, DRADVR15, GDGR}, we may further reduce the degree of computations by taking a quadrupolar approximation of the secular part due to the Earth (up to the second power of $R_E$).
In \equ{ham_sec_res_all} we neglected the
Keplerian part, since $M$ is an ignorable variable and, therefore, $L$ is constant.

As pointed out in \cite{HughesI} (see also \cite{CGPbif, DRADVR15, GDGR, aR08}), some resonances turn out to be independent on $a$, $e$, and they depend only on the inclination. The general class of resonances depending only on the inclination is characterized by the relation $k_1 \dot{\omega}+k_2 \dot{\Omega}=0$, $k_1, k_2 \in \mathbb{Z}$. From this class, the most important ones are those for which $k_1, k_2 \in \{-2,-1,0,1,2\}$. In fact, under the quadrupolar
approximation (see \eqref{Rsun_double_average} and \eqref{Rmoon_double_average}), the only possible resonances are: (i) the critical inclination resonance $\dot\omega=0$ at $63.4^{\circ}$ or $116.4^{\circ}$; (ii) the polar resonance $\dot\Omega=0$ at $90^{\circ}$; (iii) and the linear combinations: $\dot\omega+\dot\Omega=0$ at $46.4^{\circ}$ or $106.9^{\circ}$,  $-\dot\omega+\dot\Omega=0$ at $73.2^{\circ}$ or $133.6^{\circ}$,  $-2\dot\omega+\dot\Omega=0$ at $69.0^{\circ}$ or $123.9^{\circ}$, $2\dot\omega+\dot\Omega=0$ at $56.1^{\circ}$ or $111.0^{\circ}$.

Since the GNSS constellations are located close to the inclinations 56.1 and 64.3,  the most significant resonances from the practical perspective are the following ones:  $2 \dot{\omega} +\dot{\Omega}=0$  for Galileo, GPS and
BeiDou, and $\dot{\omega}=0$ for GLONASS.
Current studies (see \cite{ADRRVDM2016, RDGF2015, RDAVRD, Sanchez15} and the references therein) investigate some end-of-life disposal strategies for the GNSS constellations,  in order to avoid, in the future, the problems already faced in the LEO and GEO environments.

The resonances involving $k_1$ and $k_2$ with $|k_1|>2$ or/and $|k_2|>2$ occur at higher degree expansions of the lunar and solar disturbing functions, their influence being negligible in the MEO region.

Since  a cosine argument of $\overline{\R}_{Moon}$ could depend also on $\Omega_M$, which varies periodically, then beside the above
mentioned resonances depending only on inclinations, one also has the commensurability relations \eqref{secresmoonbis} that involve the frequency $\dot{\Omega}_M$. These resonances depend also on the eccentricity and semimajor axis. One can reformulate the problem by saying that each resonance of the classes (i), (ii) and (iii)  splits into a multiplet of resonances. This splitting phenomenon is responsible for the existence of a very complex web--like background of resonances in the phase space, which leads to a chaotic variation of the orbital elements. An analytical estimate of the location of the resonance corresponding to each component of the multiplet, as a function of eccentricity and inclination, can be obtained by using $\eqref{MomeagaOmega_var}$ (see, for example,
Figure~2 in \cite{ElyHowell} or \cite{aR15}). In \cite{CGfrontier, CGPRnote}, we have shown the web structure of resonances in the space of the actions, emphasizing how resonances overlap for various values of the semimajor axis.

Let us mention that, beside the splitting and overlapping phenomena, the dynamics of the secular resonances shows
another interesting behavior, namely the bifurcation of equilibria, for which we refer to \cite{CGfrontier, CGPbif} for further details.

\subsection{Semi--secular resonances}\label{sec:other}

According to the classification of the harmonic terms of the expansions \eqref{Rsun} and \eqref{Rgood}, we define the semi--secular resonances as follows (compare with \cite{HughesII}).
\begin{definition}
A solar semi--secular resonance occurs whenever
$$
(l-2p)\dot\omega+m\dot\Omega-(l-2h+j) \dot{M}_S=0\ , \qquad l \in \mathbb{Z}_+\,,\ m,p,h=0,1,2,...,l\,,\ j \in \mathbb{Z}.
$$
We have a lunar semi--secular resonance whenever
\beqano
(l-2p)\dot\omega+m\dot\Omega \pm [(l-2q) \dot{\omega}_M+(l-2q+r) \dot{M}_M+s\dot{\Omega}_M]=0\ ,\nonumber\\
\qquad\qquad\qquad\qquad\qquad\qquad\qquad\qquad\ l \in \mathbb{Z}_+\,,\ m,p,q,s=0,1,2,...,l\,,\ r \in \mathbb{Z}.
\eeqano
\end{definition}

By taking a quadrupolar approximation of the expansions \eqref{Rsun} and \eqref{Rgood}, namely considering $l=2$, it follows that the possible resonances have the form:
\begin{equation}\label{solar2_semi_secular_res}
\alpha \dot\omega+ \beta \dot\Omega-\gamma \dot{M}_S=0\ , \qquad \alpha \in \{\pm 2, 0\}\,, \quad\beta \in \{\pm 2, \pm 1, 0\}\,, \quad
\gamma \in \mathbb{Z}\backslash\{0\}
\end{equation}
for the Sun and
\begin{equation}\label{lunar2_semi_secular_res}
\begin{split}
& \alpha \dot\omega+ \beta \dot\Omega +\alpha_M \dot\omega_M+ \beta_M \dot\Omega_M-\gamma \dot{M}_M=0\ , \qquad \alpha\,, \alpha_M \in \{\pm 2, 0\}\,,\nonumber\\
& \hspace{7cm} \beta, \beta_M \in \{\pm 2, \pm 1, 0\}, \quad \gamma \in \mathbb{Z}\backslash\{0\}\nonumber
\end{split}
\end{equation}
for the Moon.

In the remainder of this Section, we show that such resonances are possible for relatively small values of the semimajor axis, typically in LEO and in regions close to LEO, provided the value of
the eccentricity does not exceed a threshold value. For instance,
if $e<0.4$, then a semi--secular resonance can occur only if $a<18\,611\, km$. For smaller values of the eccentricity, the bounds on the semimajor axis are smaller than $18\,611\, km$. On the contrary, if the eccentricity is large enough, then semi--secular resonances may occur for every value of the semimajor axis.

Let us discuss the case of the solar semi--secular resonances, since the lunar case can be treated in a similar way.

We shall disregard the semi--secular resonances for which $\gamma=\pm 1$, since in this case the magnitude of the resonant terms is very small. Indeed,
when $l-2h+j=\pm 1$ and $l=2$, it follows that $|j|$ is an odd number. Taking into account that the semi--secular resonant terms are of the order $\mathcal{O}(e_S^{|j|})$, and $e_S$ is small, it follows that such resonances will have a small influence on the dynamics\footnote{As a comparison, for the secular resonances \eqref{secressunbis} and \eqref{secresmoonbis}, the secular resonant terms are of the order $\mathcal{O}(1)$  in  $e_S$ and  $e_M$, respectively.}.

Therefore, we can take $|\gamma|\geq 2$. Since $M_S\simeq 1^o/day$, the relations \eqref{MomeagaOmega_var}  and \eqref{solar2_semi_secular_res} yield
$$[4.98 \alpha (5 \cos^2 i-1)-9.97 \beta \cos i] (R_E/a)^{7/2} (1-e^2)^{-2}=\gamma\,.$$
For a given value of the eccentricity, say $e=0.4$, the upper bound of the region where semi--secular resonances are possible, is obtained by taking the maximum value of the function $f(i, \alpha, \beta)=|4.98 \alpha (5 \cos^2 i-1)-9.97 \beta \cos i|$ for $i \in [0^o, 180^o]$, $ \alpha\, \in \{\pm 2, 0\}$ and $ \beta\, \in \{\pm 2, \pm 1, 0\}$, namely the value 59.78, which is obtained for $i=0^o$, $\alpha=2$ and $\beta=-2$, and the minimum value of $|\gamma|$, that is $|\gamma|=2$.
Hence, we get at most the bound $a=18\,611\, km$, which occurs when computing the maximum value of $|f|$ for $e=0.4$ and $|\gamma|=2$.
A simple computation shows that, varying  $i$ in the interval $[0^o, 180^o]$ and $e$ in the interval $[0, 0.4]$, it follows that the majority of semi--secular resonances occur most likely in LEO or nearby LEO.

\section{Milankovitch variables}\label{sec:Milankovitch}
The set of coordinates to which we refer as \emph{Milankovitch elements} (\cite{Milankovitch1939}) can be conveniently used to
describe the satellite's dynamics as shown in \cite{Rosengren2014}. This set of coordinates uses two vectorial integrals of the 2-body problem. The first vector integral is the angular momentum vector, say $\mathbf{H}$, which is assumed to be perpendicular to the instantaneous orbital plane
and it is equal to the double of the areal velocity. The second vector integral is the
Laplace-Runge-Lenz vector, say $\mathbf{b}=\mu_E \mathbf{e}$, where $\mathbf{e}$ denotes the eccentricity
vector (see also \cite{AC}).

Averaging over the mean anomaly of the particle and limiting ourselves to consider the secular Hamiltonian, the semimajor axis is constant;
we can scale $\mathbf{H}$ by the factor $\sqrt{\mu a}$ and define the scaled angular momentum as
$$
\mathbf{h}={1\over \sqrt{\mu a}} \mathbf{r}\wedge \mathbf{v}\ ,
$$
where $\mathbf{r}$ is the position vector of the particle and $\mathbf{v}$ its velocity in an
inertial frame.
We can express the eccentricity vector as
$$
\mathbf{e}={1\over \mu}\ \mathbf{v}\wedge (\mathbf{r}\wedge \mathbf{v})-
\mathbf{\hat r}\ ,
$$
$\mathbf{\hat r}$ being the unit state vector.

Denoting by $\mathbf{\overline{h}}$, $\mathbf{\overline{e}}$ the averaged vectors and setting
$\overline{h}$, $\overline{e}$ the norms of $\mathbf{\overline{h}}$, $\mathbf{\overline{e}}$, then the equations
of motion are given by
\beqa{he}
\mathbf{\dot{\bar h}}&=&\mathbf{\bar{h}}\wedge{1\over \sqrt{\mu a}}
{{\partial\bar R(\bar h,\bar e)}\over {\partial \mathbf{\bar{h}}}}+
\mathbf{\bar{e}}\wedge{1\over \sqrt{\mu a}}
{{\partial\bar R(\bar h,\bar e)}\over {\partial \mathbf{\bar{e}}}}\nonumber\\
\mathbf{\dot{\bar e}}&=&\mathbf{\bar{e}}\wedge{1\over \sqrt{\mu a}}
{{\partial\bar R(\bar h,\bar e)}\over {\partial \mathbf{\bar{h}}}}+
\mathbf{\bar{h}}\wedge{1\over \sqrt{\mu a}}
{{\partial\bar R(\bar h,\bar e)}\over {\partial \mathbf{\bar{e}}}}\ ,
\eeqa
where $\bar R$ is the average over the mean anomaly of the disturbing function
$R$ obtained as the  sum of the energy potentials due to the Earth
($V_{GEO}$), Sun ($V_{Sun}$), Moon ($V_{Moon}$) and SRP ($V_{SRP}$).
Following \cite{Rosengren2014}, the equations \equ{he} admit two
integrals: $\mathbf{\bar{h}}\cdot \mathbf{\bar{e}}$ and
$\mathbf{\bar{h}}\cdot \mathbf{\bar{h}}+\mathbf{\bar{e}}\cdot \mathbf{\bar{e}}$.
To get physically meaningful solutions, one needs to consider the motion on the manifold
restricted to $\mathbf{\bar{h}}\cdot \mathbf{\bar{e}}=0$ and
$\mathbf{\bar{h}}\cdot \mathbf{\bar{h}}+\mathbf{\bar{e}}\cdot \mathbf{\bar{e}}=1$
(\cite{Tremaine2009}).

We now list the averaged components of the potential in terms of the Milankovitch elements (see \cite{Rosengren2013}).

Concerning the geopotential $V_{GEO}$ reduced to its main contribution through the $J_2$ term, we have
that the averaged potential is given by
$$
\bar V_{GEO}={{n C_{20}}\over {4a^2 h^3}}\ \left[1-3(\mathbf{\hat p}\cdot \mathbf{\hat h})^2\right]\ ,
$$
where $n$ is the mean motion $n=(\mu_E/a^3)^{1\over 2}$,
$\mathbf{\hat p}$ is the unit vector along the direction of the Earth's maximum axis of inertia.
For the solar potential $V_{Sun}$ and the lunar potential $V_{Moon}$,  we have that,
under the quadrupolar approximation, their averages can be written as
$$
\bar V_{Sun}={{3\mu_S}\over {4n d_S^3}}\ \left[5(\mathbf{\hat d_S}\cdot\mathbf{e})^2-
(\mathbf{\hat d_S}\cdot\mathbf{h})^2-2e^2\right]\ ,
$$
$$
\bar V_{Moon}={{3\mu_M}\over {4n d_M^3}}\ \left[5(\mathbf{\hat d_M}\cdot\mathbf{e})^2-
(\mathbf{\hat d_M}\cdot\mathbf{h})^2-2e^2\right]\ ,
$$
where $\mathbf{\hat d_S}$ is the unit vector of the Sun with respect to the Earth, and $\mathbf{\hat d_M}$ the one to the Moon with respect to Earth,
the quantities $D_S$, $d_M$, $e$ represent the norms of $\mathbf{d_S}$, $\mathbf{d_M}$, $\mathbf{e}$.
Finally, the averaged potential for SRP is
$$
\bar V_{SRP}={3\over 2}\ \sqrt{a\over \mu}\ {\beta\over {d_S^2}}\ \mathbf{\hat d_S}\cdot\mathbf{e}\ ,
$$
where $\beta=(1+\rho)A/m P_\Phi$ with $\rho$ the reflectance, $A/m$ the area-to-mass ratio and $P_\Phi$ the solar radiation constant.
In conclusion, the secular equations in terms of Milankovitch elements are given by \equ{he} with
$\bar R=\bar V_{GEO}+\bar V_{Sun}+\bar V_{Moon}+\bar V_{SRP}$.

\section{Epicyclic variables}\label{sec:epicyclic}

A different approach to model an Earth-orbiting particle can be given in terms of the so-called \sl epicyclic variables \rm (see \equ{epivar} below).
The Hamiltonian formulation in terms of the epicyclic variables turns out to be convenient, since
i) it puts the Hamiltonian in action-angle variables, and ii) it simplifies the algebra compared to regular expansions in elements made in most analytical studies. We recall here a model developed in \cite{Gachet2016},
particularly apt to study the GEO region, containing all major perturbations: the geopotential including the zonal
coefficient $J_2$ and the tesseral $J_{22}$ terms,
the solar potential up to order 2 in the ratio of the geocentric distances to
the particle and to the third body, the Moon's one up to order 4, due to its proximity,
and the solar radiation pressure using the cannonball model and neglecting Earth shadows. We follow here the treatment of \cite{Gachet2016},
where a step-by-step approach is taken.

We consider an Earth-centered inertial reference frame, whose $z$-axis is aligned with the rotation axis of the Earth, and whose $x$-axis points for instance towards the mean equinox on JAN 1 2000 at noon (EME2000 reference frame), the $y$-axis completing the right-handed frame. We attach to this frame the classical cylindrical coordinates $(\rho,\Phi,z)$.
The Hamiltonian of the system is then:
$$
H(p_\rho,p_\Phi,p_z,\rho,\Phi,z,t)=
\frac{{p_\rho}^2}{2}+\frac{{p_\Phi}^2}{2\rho^2}+\frac{p_z^2}{2}+V(\rho,\Phi,z,t)\ ,
$$
where
$$
p_\rho=\dot{\rho}\ ,\qquad p_\Phi=\rho^2\dot{\Phi}\ , \qquad p_z=\dot{z}\ ,  \nonumber
$$
and $V$ represents the potential derived from all forces accounted for in the model.
We have:
$$
V=V_{GEO}+V_{Moon}+V_{Sun}+V_{SRP}\ ,
$$
where $V_{GEO}$ is the geopotential, $V_{Moon}$, $V_{Sun}$ the gravitational perturbation potentials of the Moon and Sun respectively, and $V_{SRP}$ the solar radiation pressure potential. The dependence on time of $V$ comes from the position of the
Moon and Sun; to make the system autonomous and to make explicit the intrinsic frequencies of the system, we extend the phase space by adding 5 degrees of freedom. To this end we introduce:
$$
\varphi_E=\Omega_E t\ , \quad \varphi_M=\Omega_M t\ , \quad \varphi_{M_a}=\Omega_{M_a} t\ ,
\quad \varphi_{M_p}=\Omega_{M_p} t\ , \quad \varphi_{M_s}=\Omega_{M_s} t
$$
with
\begin{equation}
\begin{aligned}
\Omega_E&=131850^\circ \ \text{yr}^{-1}\nonumber\\
\Omega_M&=359.99049^\circ \ \text{yr}^{-1}\nonumber\\
\Omega_{M_a}&=4771.9886753^\circ \ \text{yr}^{-1}\nonumber\\
\Omega_{M_p}&=40.6901335^\circ \ \text{yr}^{-1}\nonumber\\
\Omega_{M_s}&=19.3413784^\circ \ \text{yr}^{-1}.\nonumber
\end{aligned}
\end{equation}
In the above formulae,
$\Omega_E$ denotes the Earth's sidereal rotation rate about its axis in inertial space with an associated period of about one day, $\Omega_M$ the rotation rate of the Earth around the Sun with an associated period of about about one year, and $\Omega_{M_a}$, $\Omega_{M_p}$, $\Omega_{M_s}$ are associated to the Moon's motion. More precisely $\Omega_{M_a}$ is the monthly rotation rate of the Moon, here specifically  equal to the anomalistic month, $\Omega_{M_p}$ is linked to the precession of its perigee with an associated period of 8.85 years, and finally $\Omega_{M_s}$ is linked to the precession of its node with an associated period of 18.6 years.
These new angles $(\varphi_E,\varphi_M,\varphi_{M_a},\varphi_{M_p},\varphi_{M_s})$ are associated with conjugate momentum variables, the so-called
\sl dummy action \rm variables $(I_E,J_M,J_{M_a},J_{M_p},J_{M_s})$. The extended Hamiltonian now reads
\begin{equation}\label{eq:HamSphExt}
\begin{aligned}
&H(p_\rho,p_\Phi,p_z,I_E,J_M,J_{M_a},J_{M_p},J_{M_s},\rho,\Phi,z,\varphi_E,\varphi_M,\varphi_{M_a},\varphi_{M_p},\varphi_{M_s})=\\
&=\frac{{p_\rho}^2}{2}+\frac{{p_\Phi}^2}{2\rho^2}+\frac{p_z^2}{2}
+V(\rho,\Phi,z,\varphi_E,\varphi_M,\varphi_{M_a},\varphi_{M_p},\varphi_{M_s})\\
&+\Omega_E I_E+\Omega_M J_M+\Omega_{M_a} J_{M_a}
+\Omega_{M_p} J_{M_p}+\Omega_{M_s} J_{M_s}~~.
\end{aligned}
\end{equation}
The final step to render the Hamiltonian autonomous is the introduction of the angle $\varphi=\Phi-\Omega_E t$,
which corresponds to the longitude in an Earth fixed reference frame, proving convenient since the geopotential depends on this quantity.
Through the canonical transformation $\Phi=\varphi+\varphi_E$, $p_\Phi=p_\varphi$, $I_E=J_E-p_\varphi$,
the Hamiltonian \equ{eq:HamSphExt} takes the form
\begin{equation}\label{eq:HamSphe}
\begin{aligned}
&H\equiv H(p_\rho,p_\Phi,p_z,J_E,J_M,J_{M_a},J_{M_p},J_{M_s},\rho,\varphi,z,\varphi_E,\varphi_M,\varphi_{M_a},\varphi_{M_p},\varphi_{M_s})=\\
&=\frac{{p_\rho}^2}{2}+\frac{{p_\varphi}^2}{2\rho^2}+\frac{p_z^2}{2} - \Omega_E p_\varphi
+V(\rho,\varphi,z,\varphi_E,\varphi_M,\varphi_{M_a},\varphi_{M_p},\varphi_{M_s})\\
&+\Omega_E J_E+\Omega_M J_M+\Omega_{M_a} J_{M_a}
+\Omega_{M_p} J_{M_p}+\Omega_{M_s} J_{M_s}~~.
\end{aligned}
\end{equation}

Under the quadrupolar assumptions mentioned above, we have that $V_{GEO}$ is given by
\begin{equation}
\begin{aligned}
V_{GEO}&=-\frac{\mu_E}{\sqrt{\rho^2+z^2}}+\frac{\sqrt{5} \bar C_{2,0} \mu_E R_E^2}{2(\rho^2+z^2)^{3/2}}-\frac{3 \sqrt{5}
\bar C_{2,0} \mu_E R_E^2 z^2}{2(\rho^2+z^2)^{5/2}}\nonumber\\
&-\frac{\sqrt{15} \mu_E R_E^2}{2(\rho^2+z^2)^{3/2}}\left(1-\frac{z^2}{\rho^2+z^2}\right)\left(\bar C_{2,2}
\cos (2 \varphi ) +\bar S_{2,2} \sin (2 \varphi) \right)\ ,\nonumber
\end{aligned}
\end{equation}
where $\bar C_{nm}$, $\bar S_{nm}$ are the normalized spherical
harmonic coefficients, defined as
\beqano
\bar C_{nm}&=&\sqrt{{{(n+m)!}\over{2(2n+1)(n-m)!}}}\ C_{nm}\ ,\quad \bar
C_{n0}={1\over {\sqrt{2n+1}}}C_{n0}\ ,\nonumber\\
\bar S_{nm}&=&\sqrt{{{(n+m)!}\over{2(2n+1)(n-m)!}}}\ S_{nm}\ ,\ \ m>0\ .
\eeqano

We recall that the lunisolar perturbations are described by the following potentials:
$$
V_{Sun}=-\G m_S \left( \frac{1}{|\mathbf{r}-\mathbf{r_S}|} - \frac{\mathbf{r}\cdot\mathbf{r_S}}{|\mathbf{r_S}|^3} \right)
$$
and
$$
V_{Moon}=-\G m_{M} \left( \frac{1}{|\mathbf{r}-\mathbf{r_{M}}|} - \frac{\mathbf{r}\cdot\mathbf{r_{M}}}{|\mathbf{r_{M}}|^3} \right)
$$
with the particle's state vector given by
$$
\mathbf{r}=\left(
\begin{array}{c}
\rho \cos \Phi \\
\rho \sin \Phi\\
z
\end{array} \right)
=\left(
\begin{array}{c}
\rho \cos (\varphi+\varphi_E) \\
\rho \sin (\varphi+\varphi_E)\\
z
\end{array} \right)\ ,
$$
while $\mathbf{r_S}$ and $\mathbf{r_M}$ are the Sun and Moon state vectors, respectively.
Their expressions can be found in \cite{MG} and depend on $(\varphi_M,\varphi_{M_a},\varphi_{M_p},\varphi_{M_s})$.
We refer to \cite{Gachet2016} for full a complete description of the equations of motion.

The solar radiation pressure energy potential under the cannonball approximation reads as
$$
V_{SRP}=C_rP_r\ a_S^2\ \frac{A}{m}\ \frac{1}{|\mathbf{r}-\mathbf{r_S}|}
$$
with $C_r$ the reflectivity coefficient, $ P_r=4.56\times 10^{-6}$~Nm\textsuperscript{-2} is the radiation pressure for
an object located at $a_S=1$ AU, and $A/m$ the area-to-mass ratio.
We point out that it is crucial to consider the Sun moving on an inclined ellipse for the SRP potential, otherwise, as \cite{VLA} remarks, having a fixed Sun-Earth distance in the estimation of SRP can induce spurious long-period terms in eccentricity and inclination evolution.

To introduce epicyclic variables to study the motion at GEO, one needs to define the geostationary radius. To this end we isolate the axisymmetric
part of the geopotential, that we write as
$$
V_{GEO_0}(\rho,z)=-\frac{\mu_E}{\sqrt{\rho^2+z^2}}+\frac{\sqrt{5} \bar C_{2,0} \mu_E R_E^2}{2(\rho^2+z^2)^{3/2}}-\frac{3 \sqrt{5} \bar C_{2,0} \mu_E R_E^2 z^2}{2(\rho^2+z^2)^{5/2}}\ .
$$
The angular velocity of an equatorial circular orbit at the distance $\rho$ is given by
$$
W(\rho)=\sqrt{\frac{1}{\rho}\frac{dV_{GEO_0}(\rho,z)}{d\rho}\Bigg|_{z=0}}\ .
$$
The radius $\rho_c$ at which $W(\rho_c)=\Omega_E$ is the geostationary radius.
The angular momentum per unit mass of a particle in circular orbit at the geostationary radius is equal to $p_c=\Omega_E \rho_c^2$. We then call
\sl effective potential \rm the quantity
$$
V_{GEO_{eff}}=\frac{p_c^2}{2\rho^2}+V_{GEO_0}(\rho,z)~~.
$$
The effective potential describes the epicyclic oscillations of particles in nearly circular orbits under the axisymmetric potential $V_{GEO_0}(\rho,z)$
with (preserved) value of the $z$-component of the angular momentum $p_\varphi=p_c$.
The radial and vertical epicyclic frequencies are then $\kappa_\rho$ and $\kappa_z$, respectively, with
$$
\kappa_\rho=\sqrt{\frac{d^2V_{GEO_{eff}}}{d\rho^2}}\ ,\qquad
\kappa_z=\sqrt{\frac{d^2V_{GEO_{eff}}}{dz^2}}\ .
$$
This definition stems from the fact that setting $\rho=\rho_c+\delta\rho$, and expanding $V_{GEO_{eff}}$ up to terms of second degree in $\delta\rho$ and $z$, we have
$$
V_{GEO_{eff}}=const + {1\over 2}\kappa_\rho^2\delta\rho^2 + {1\over 2}\kappa_z^2 z^2 +\ldots
$$

In a neighborhood of the geostationary radius, it is natural to define the displacement $\delta\rho=\rho-r_c$,
$J_\varphi=p_\varphi-p_c$. Expanding the Hamiltonian in $\delta\rho$ and $J_\varphi$, we obtain from (\ref{eq:HamSphe}):
\begin{equation}
\begin{aligned}
H&=\left({p_\rho^2\over 2} + {p_z^2\over 2} +{1\over 2}\kappa^2\delta\rho^2+{1\over 2}\kappa_z^2z^2\right)
+\Omega_E J_E + \Omega_M J_M
+ \Omega_{M_a} J_{M_a} + \Omega_{M_p} J_{M_p} + \Omega_{M_s} J_{M_s} \nonumber\\
&+ H_{pert}(J_\varphi,\delta \rho,\varphi,z,\varphi_E,\varphi_M,\varphi_{M_a},\varphi_{M_p},\varphi_{M_s})\nonumber
\end{aligned}
\end{equation}
with $H_{pert}$ a polynomial in $\delta\rho$, $z$, up to second degree in $J_\varphi^2$, and trigonometric in all the angular variables.

We finally introduce the epicyclic action-angle variables $(J_\rho,\varphi_\rho)$ and $(J_z,\varphi_z)$ defined as
\begin{equation}
\label{epivar}
\begin{aligned}
\delta \rho &= \sqrt{\frac{2 J_\rho}{\kappa}} \sin \left(\varphi_\rho\right)\ ,\qquad
z = \sqrt{\frac{2 J_z}{\kappa_z}} \sin \left(\varphi_z\right)\ ,\\
p_{\rho } &= \sqrt{2 \kappa J_\rho} \cos \left(\varphi_\rho\right)\ ,\qquad
p_z = \sqrt{2 J_z \kappa_z} \cos \left(\varphi_z\right)\ ,
\end{aligned}
\end{equation}
which lead to the following Hamiltonian
\begin{equation}
\label{eq:Hnew}
\begin{aligned}
&H(J_\rho,J_\varphi,J_z,J_E,J_{M},J_{M_a},J_{M_p},J_{M_s},\varphi_\rho,\varphi,\varphi_z,\varphi_E,\varphi_{M},\varphi_{M_a},\varphi_{M_p},\varphi_{M_s})\\
&=\kappa J_\rho + \kappa_zJ_z +\Omega_E J_E + \Omega_M J_M
+ \Omega_{M_a} J_{M_a} + \Omega_{M_p} J_{M_p} + \Omega_{M_s} J_{M_s}+\text{h.o.t}\ ,
\end{aligned}
\end{equation}
where h.o.t. denotes terms of order higher than 2 in the actions. From the Hamiltonian
\equ{eq:Hnew} we understand
the convenience of the epicyclic variables as action-angle variables, since at the zeroth order, we have the integrable part depending just on the actions,
coupled with the associated frequencies of the angles.
One can then study the dynamics by applying the method of normal forms by Lie Series as done in \cite{Gachet2016}
(see \cite{YX2, YX3, YX1} for further references).

\section{Onset of chaos in the conservative regime}\label{sec:chaos}
The determination of the regular and chaotic behavior of space debris is nowadays of seminal
importance, since the different character of the dynamics might strongly contribute whether to
place a debris in a stable region or rather move it toward a chaotic zone. In particular,
inserting a space debris along the unstable manifold of an hyperbolic equilibrium point
might allow to move the debris without too much effort toward convenient regions, even possibly the
graveyard zones.
Hence, the transition from a regular to a chaotic motion can be used to move the debris
within different regions, possibly paving the way to the design of disposal orbits. This is a focus
topic, which certainly deserves dedicated studies. However, this analysis cannot be performed without
an accurate knowledge of the mechanisms leading to chaos, which will be summarized in
Sections~\ref{sec:overlapping}-\ref{sec:lunisolar}.

In the past years much effort has been devoted to understand which are the
most important factors which contribute to the onset of chaos. Such analysis strongly depends
on the region where the debris is located, since - as we already mentioned - in LEO the dissipative
atmospheric drag plays a special role, in MEO lunisolar secular resonances are particularly
relevant, in GEO the effects of Sun, Moon and SRP strongly affect the dynamics. On the other hand,
the analysis depends also on the scale of time and the orbital elements emphasized, since  the overlapping of tesseral resonances leads to a chaotic variation of the semimajor axis on a relatively short time (tens to hundreds
of days), while the eccentricity varies chaotically, as effect of the overlapping of secular resonances, on a much longer (secular) timescale.

In this Section we review some of the main effects which contribute to the onset of chaos.
The list is not intended to be exhaustive, but rather to give an idea of how chaos is generated
or, maybe, could be even artificially induced.

\subsection{Overlapping of tesseral resonances}\label{sec:overlapping}
As effect of the secular part $\H^{sec}$, which is dominated by a term of the order of
magnitude of $J_2$, the frequencies $\dot{\omega}$ and $\dot{\Omega}$ are not zero, but rather they may be computed as a function of eccentricity and inclination  by using the relations \eqref{MomeagaOmega_var}. Since
the geopotential is a sum of trigonometric terms depending on the angle
$\Psi_{nmpq}=(n-2p)\omega+(n-2p-q)M+m(\Omega-\theta)$, as noted in \cite{CGmajor, CGminor},
for a specific resonance and for different values of the indexes, the angles $\Psi_{nmpq}$
are stationary at different locations, thus giving rise to a multiplet of resonances in which
each resonance is split. Taking advantage from the pendulum-like structure associated to each term of the expansion, we can estimate the amplitude corresponding to
the different components of the multiplet.
For a $j:\ell$ gravitational resonance, by retaining the Keplerian part, the secular part and the resonant term
corresponding to  the $q$--th component of the multiplet, we obtain the Hamiltonian function
\begin{equation}\label{ham_tess_res_comp_mutiplet}
\H_{q}^{j:\ell}=-\frac{\mu_E^2}{2L^2} + \H^{sec}+\A_q\,  cs\Bigl(\ell M +j (\Omega-\theta)+(\ell+q) \omega\Bigr)\,,
\end{equation}
where $\A_q=\A_q(L,G,H)$ is an explicit function of the actions and $cs$ can be either cosine or sine, as in \eqref{H21}.

\begin{figure}[hpt]
\centering
\vglue0.1cm
\hglue0.1cm
\includegraphics[width=6truecm,height=4truecm]{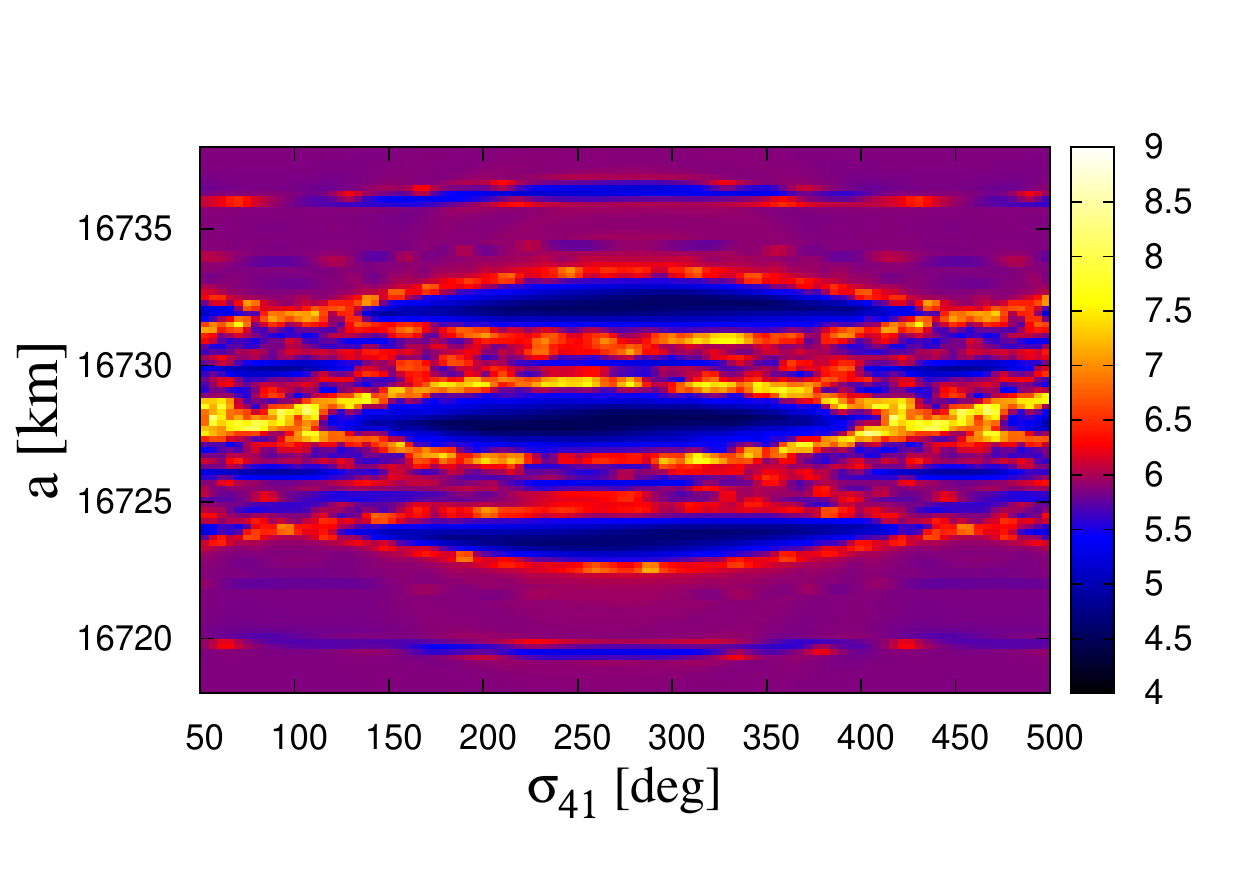}
\includegraphics[width=6truecm,height=4truecm]{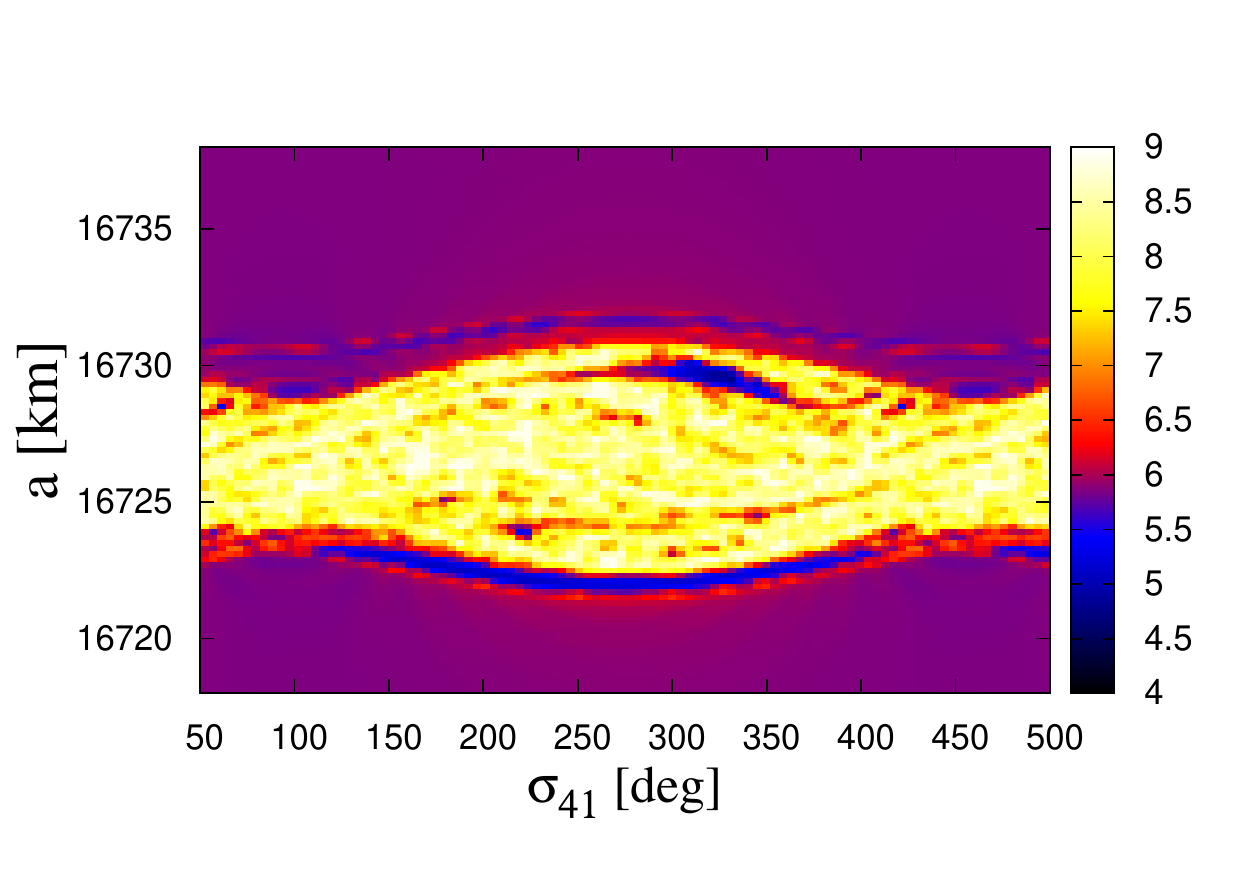}
\vglue0.5cm
\caption{
Splitting and superposition of resonances: FLI for the 4:1 resonance for $e(0)=0.3$,  $\omega(0)=0^o$,
$\Omega(0)=0^o$: $i(0)=32^o$ (left);  $i(0)=47^o$ (right). $\sigma_{41}$ is the resonant angle (compare with \cite{CGminor}) and
$a$ is the semimajor axis.}
\label{fig:split}
\end{figure}

Normalizing the units such that $\dot{\theta}=1$, then from the resonance relation and Kepler's third law, we obtain that the resonant value of the action $L$ is $L_{res}=\Bigl({{\ell\mu_E^2}\over {j}}\Bigr)^{1\over 3}$ .
Expanding \equ{ham_tess_res_comp_mutiplet} around $L_{res}$ up to second order, one is led to the pendulum--like Hamiltonian:
$$
\mathcal{H}_q^{j:\ell}=\alpha(L-L_{res})-\beta (L-L_{res})^2 +\A_q(L_{res}, G,H)\ cs\Bigl(\ell M +j (\Omega-\theta)+(\ell+q) \omega\Bigr)\ ,
$$
where
\beqano
\alpha&\equiv&{{\mu_E^2}\over {L_{res}^3}}+{{\partial \H^{sec}}\over {\partial L}}(L_{res},G,H,\omega, \Omega)\nonumber\\
\beta&\equiv& {{3\mu_E^2}\over {2L_{res}^4}}-{1\over 2}
{{\partial^2 \H^{sec}}\over {\partial L^2}}(L_{res},G,H,\omega, \Omega)\ .
\eeqano
As shown in \cite{CGminor}, the resonant island associated to the $q$--th component of the $j:\ell$ resonance has the semi-amplitude
$\Delta a_q$ given by
$$ \Delta a_q={1\over \mu_E}\ \Bigl({{2\A_q}\over \beta}+2L_{res}\
 \sqrt{{{2\A_q}\over \beta}}\Bigr)\ .
$$
On the basis of the above formula, we proceed to measure the amplitude of the resonant island associated to each component of the multiplet.
When such width is larger than the distance between nearby resonances, then we have a \sl splitting \rm
phenomenon, otherwise we have a \sl superposition \rm of resonances with a consequent onset of chaotic
motions. An example of splitting and superposition of resonances is given in Figure~\ref{fig:split}, where the Fast Lyapunov Indicator (see the Appendix for details) is computed for discriminating between regular and chaotic motions.

We remark that the onset of chaos due to overlapping of resonances can also be generated by changing the
orbital elements, most notably the eccentricity, the inclination, the argument of perigee and the longitude of the
ascending node.

\subsection{High area-to-mass ratio objects}\label{sec:Am}

\begin{figure}[h]
\centering
\vglue0.1cm
\hglue0.1cm
\includegraphics[width=6truecm,height=5truecm]{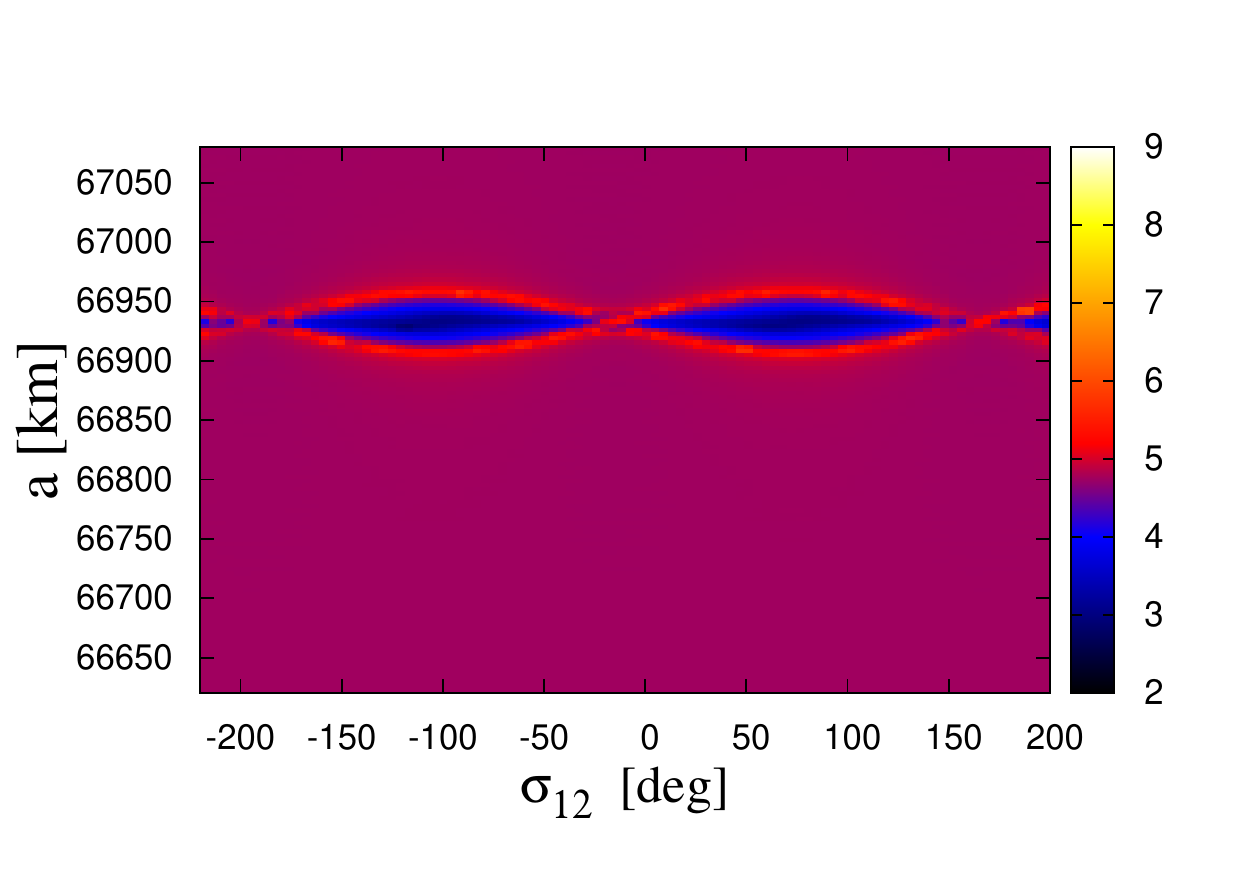}
\includegraphics[width=6truecm,height=5truecm]{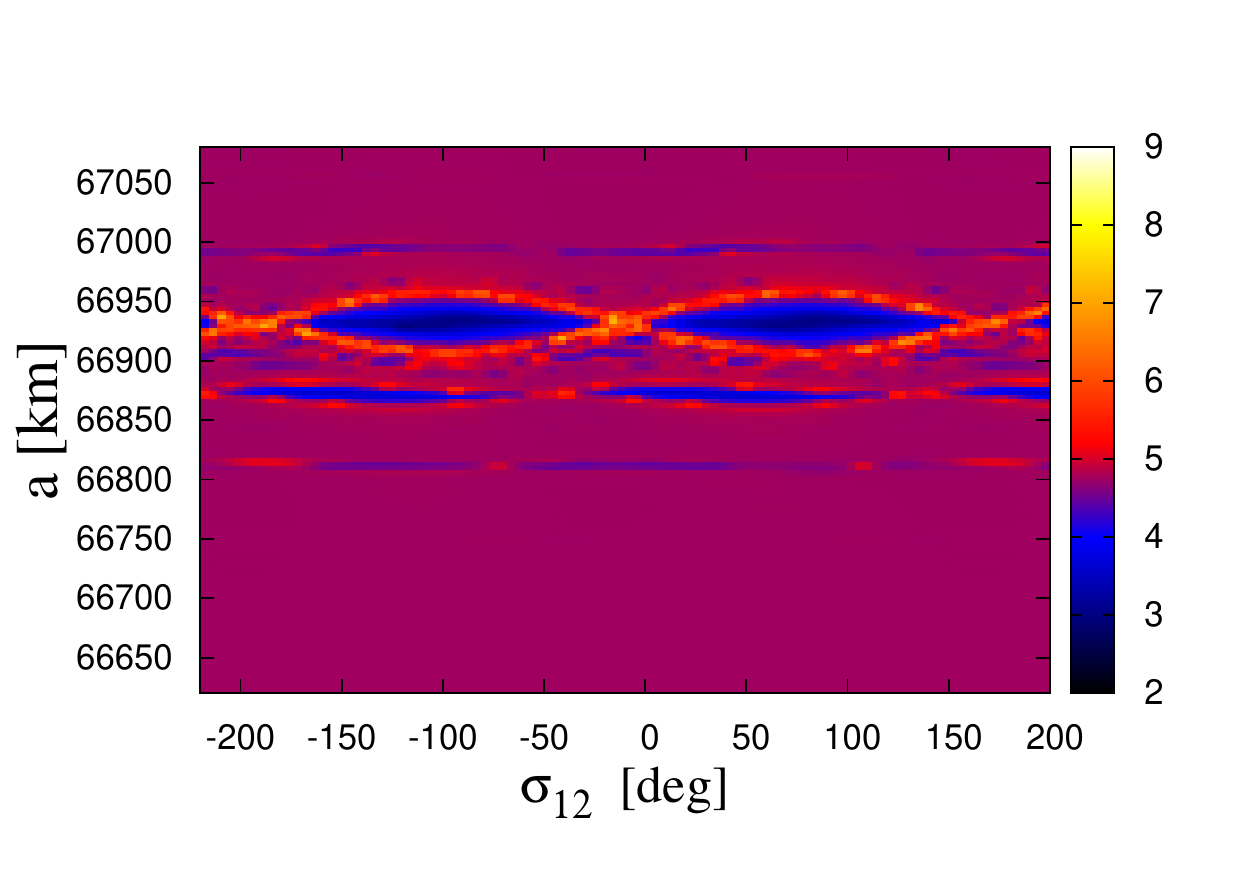}\\
\vglue-0.6cm
\includegraphics[width=6truecm,height=5truecm]{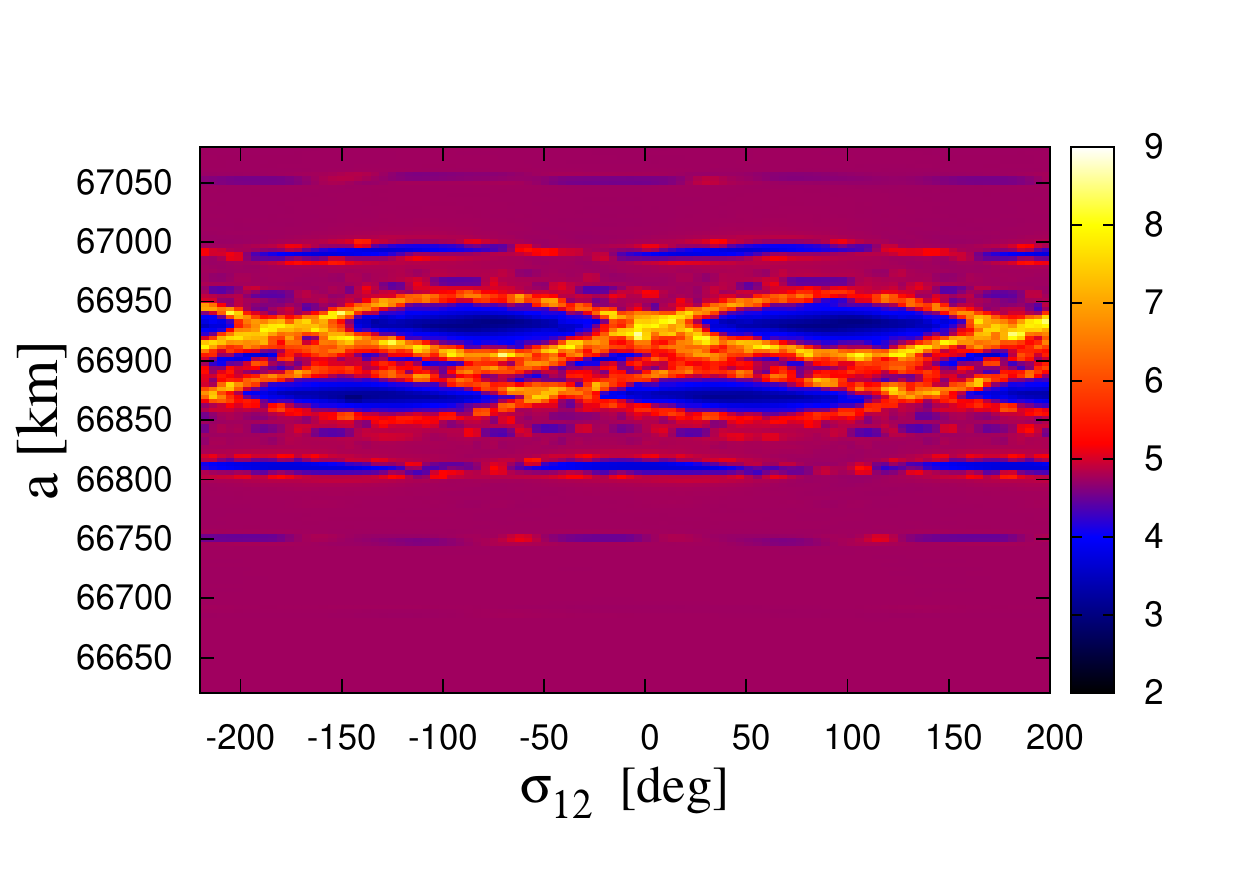}
\includegraphics[width=6truecm,height=5truecm]{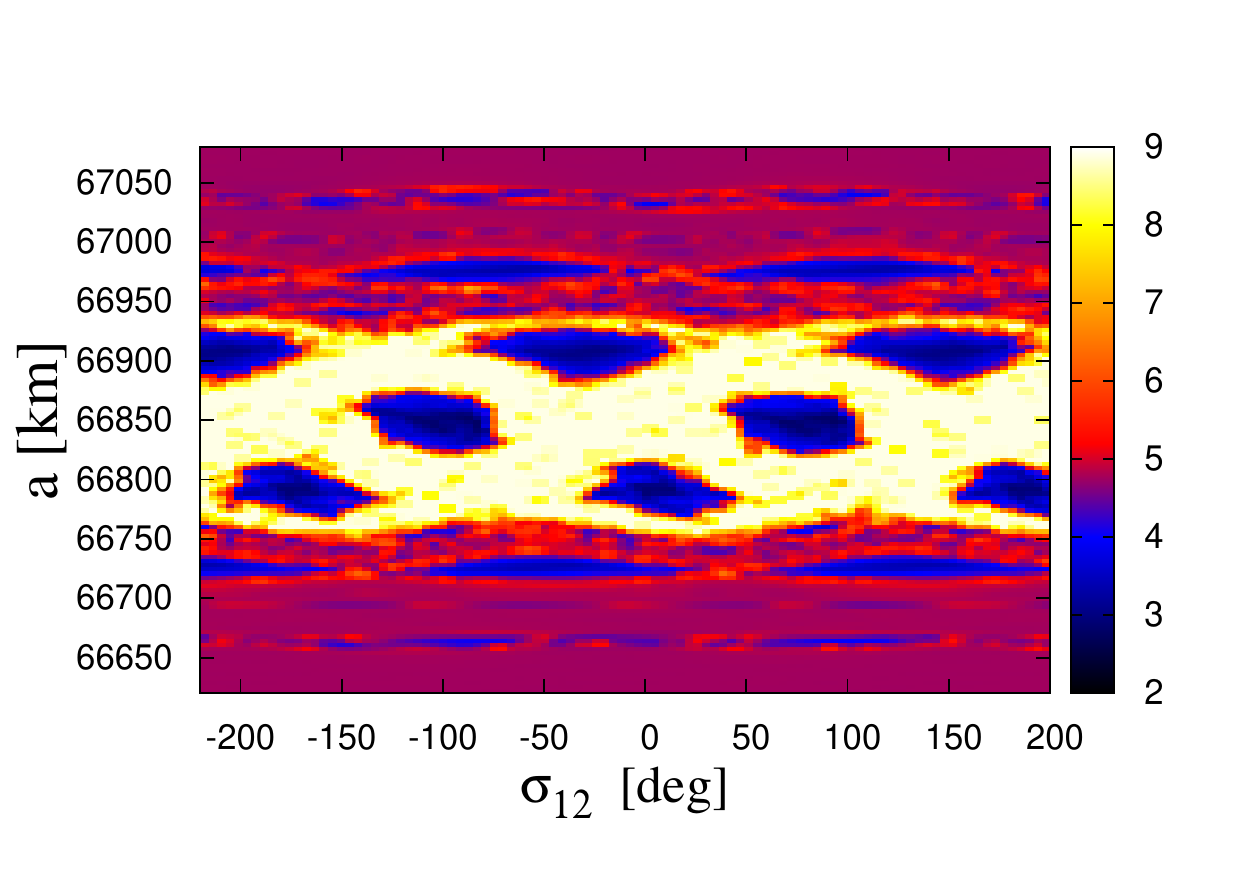}
\vglue0.4cm
\caption{FLI for the 1:2 resonance, under the effects of the geopotential and SRP, for $i(0)=0^o$, $e(0)=0.25$, $\omega(0)=0^o$, $\Omega(0)=0^o$:
 $A/m=0\, [m^2/kg]$ (top left); $A/m=1 \, [m^2/kg]$ (top right);
  $A/m=5\, [m^2/kg]$ (bottom left); $A/m=20\, [m^2/kg]$ (bottom right). $\sigma_{12}$ is the resonant
  angle (compare with \cite{CGext}) and $a$ is the semimajor axis.}
\label{Am_res12}
\end{figure}

Objects with high area-to-mass ratios have been discovered in the early
2000s (\cite{XX1}); they exhibit a peculiar behavior, and are suspected to
come from thermal insulation layers (\cite{XX2}).

We can write the potential associated to the SRP as
\beqa{VSRP}
V_{SRP}&=&C_rP_ra_S^2\ {A\over m}\ {1\over {|\mathbf{r}-\mathbf{r}_S|}}\nonumber\\
&=&C_rP_ra_S^2\ {A\over m}\ {1\over r_S}\ \sum_{j=1}^\infty ({r\over r_S})^j\ P_j(\cos Q)\ ,
\eeqa
where we denoted by $Q$ the angle between the Sun and the geocentric radius of the debris.
The position of the Sun is taken from \cite{MG}, and normalized with respect to the geostationary distance $a_{GEO}=42\,164.17 \ km$ as well as with a unit of time $\tau$ chosen such that the period of
the Earth's rotation
becomes equal to $2 \pi$. Next, we compute the expansion of \equ{VSRP} up to
third order in the Legendre polynomials, we neglect terms with coefficients less than
a specific error, and we average over the mean anomaly. The resulting expression of the approximate form
of $V_{SRP}$ is the following
(\cite{CGext}):
\beqano
V_{SRP}^{app}&=&-a\ e\ {A\over m}\ \Big(
-4.838\,10^{-7} \sin (-0.00546061\ \theta+\omega )\nonumber\\
&-&4.836\, 10^{-7}\sin (-0.00546061\ \theta+\omega  )-0.000028751\sin (-0.0027303\ \theta+\omega  )\nonumber\\
&+&1.239\,10^{-6}\sin (0.0027303\ \theta+\omega)+5.425\,10^{-6} \cos (-0.0027303\ \theta+\omega )\nonumber\\
&+&1.141\,10^{-7}\cos (\omega +0.0027303\ \theta )\Big)\ .
\eeqano

We report in Figure~\ref{Am_res12} the FLI plots for the 1:2 resonance, under the effects of the geopotential and SRP
for different values of $A/m$. Increasing the area-to-mass ratio, one gets a web of resonances which give rise to a large chaotic region covering an area of several hundreds kilometers.
Extensive studies related to the dynamics of high area-to-mass ratio geosynchronous space debris may be found in various
papers (see \cite{LDV, VLA, VL, VLD, VDLC} and references therein).
We mention that the long--term evolution of space debris under various effects, including the solar radiation pressure, was investigated in \cite{CPL2015}.

\subsection{Lunisolar secular resonances}\label{sec:lunisolar}

\begin{figure}[hpt]
\centering
\vglue0.1cm
\hglue0.1cm
\includegraphics[width=5truecm,height=4truecm]{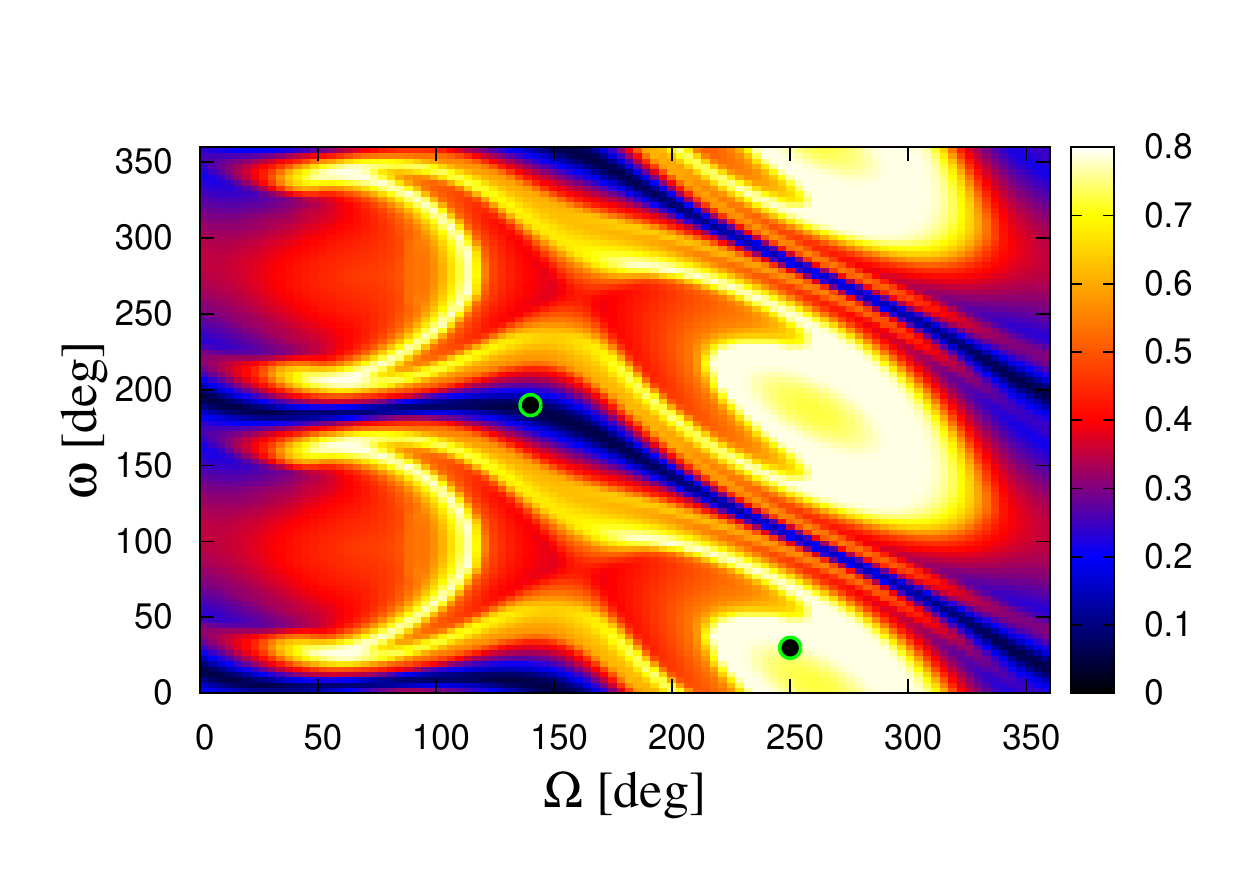}
\includegraphics[width=5truecm,height=4truecm]{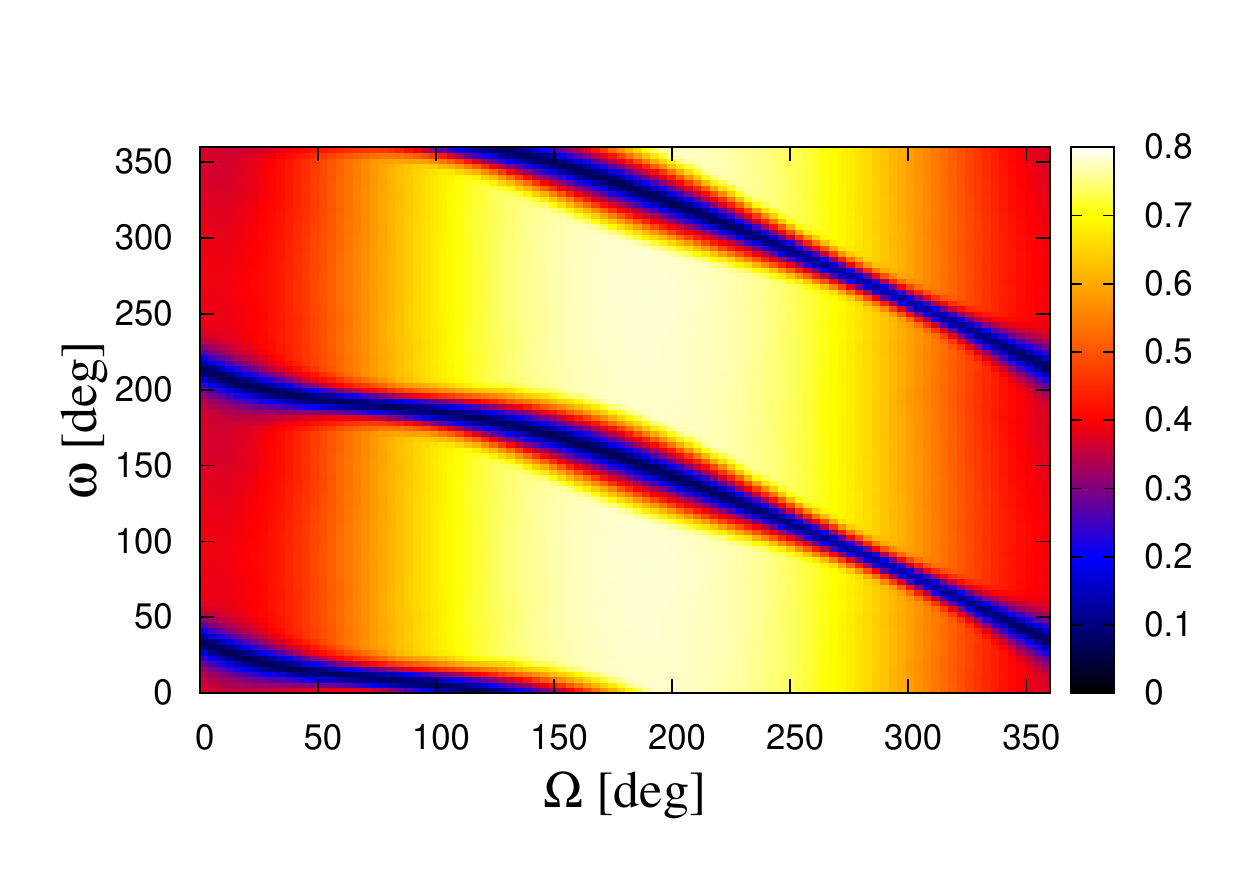}
\includegraphics[width=5truecm,height=3.4truecm]{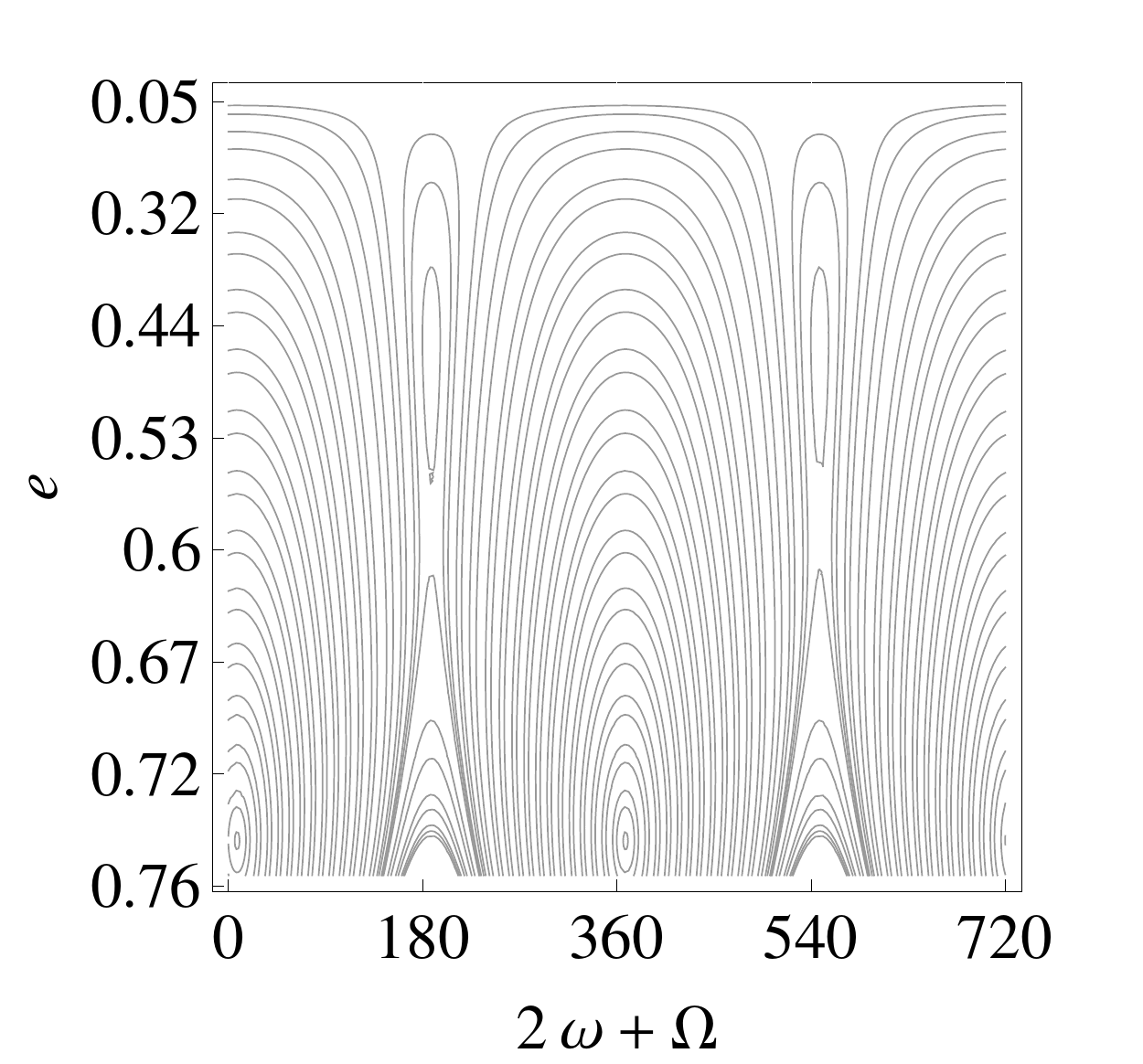}
\vglue0.5cm
\caption{
The effect of the resonance $2\dot{\omega}+\dot{\Omega}=0$. Left panel: the maximum eccentricity reached
in 200 years (color bar), as a function of the initial longitude of the ascending node $\Omega$ and the argument
of perigee $\omega$. The semimajor axis is $a=26\,520\, km$ and the initial conditions are $e(0)=0.05122$
and $i(0)=56^o$ at the initial Epoch J2000 (January 1, 2000, 12:00 GMT). The green--black circles
represent the orbits analyzed in Figure~\ref{fig:two_orbits}. Middle panel: same conditions as for
the left panel, but  $\Omega_M$ is considered constant. Right panel: bifurcation of equilibria,
as shown by the one--degree--of--freedom toy model obtained from the Hamiltonian
\eqref{ham_sec_res_all}, after passing through a canonical transformation to the resonant variables,
averaging the resulting Hamiltonian over the non--resonant angle and taking $\Omega_M$ constant
(see \cite{CGfrontier}). The phase space portrait is obtained for the same value $a=26\,520\, km$ of the
semimajor axis as for the left and middle panels.}
\label{fig:lunisolar}
\end{figure}

\begin{figure}[hpt]
\centering
\vglue0.1cm
\hglue0.1cm
\includegraphics[width=5truecm,height=4truecm]{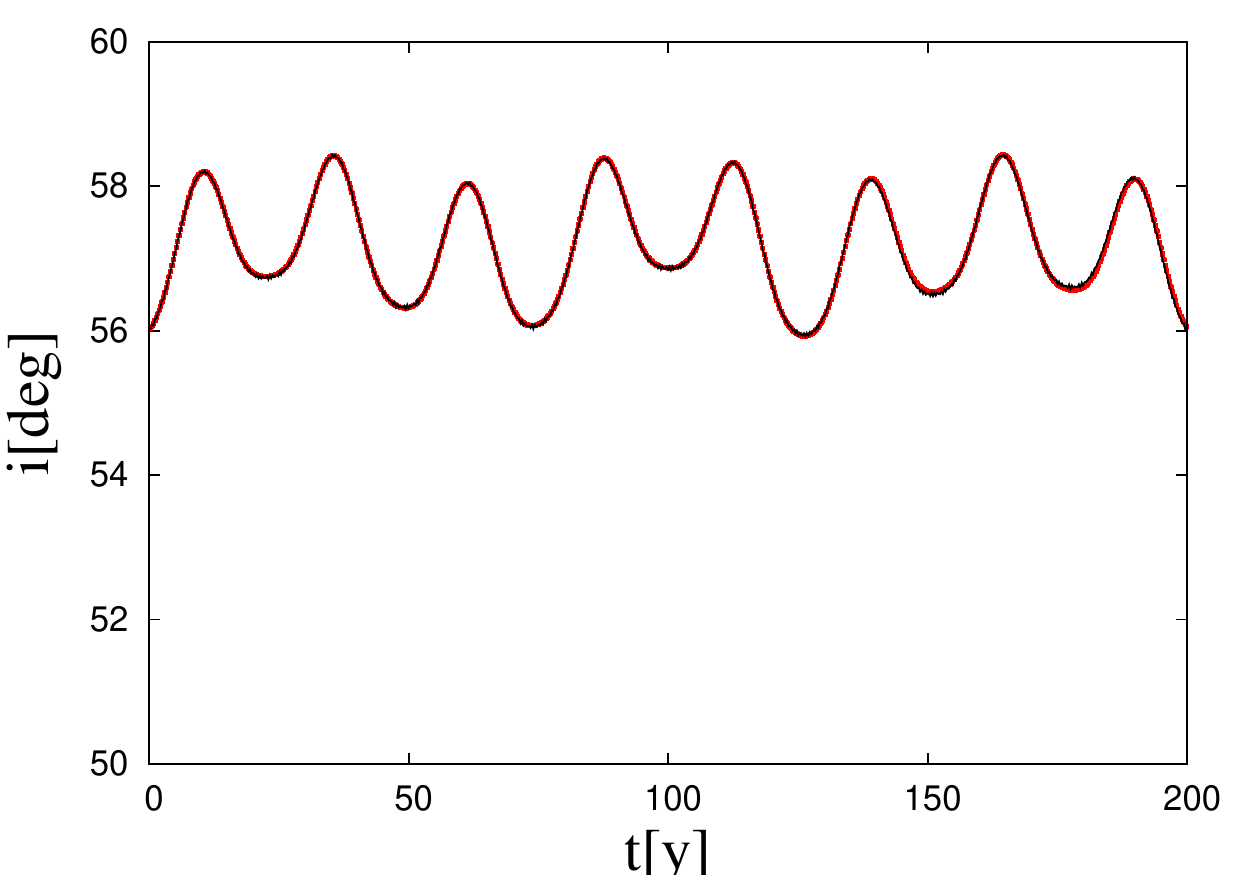}
\includegraphics[width=5truecm,height=4truecm]{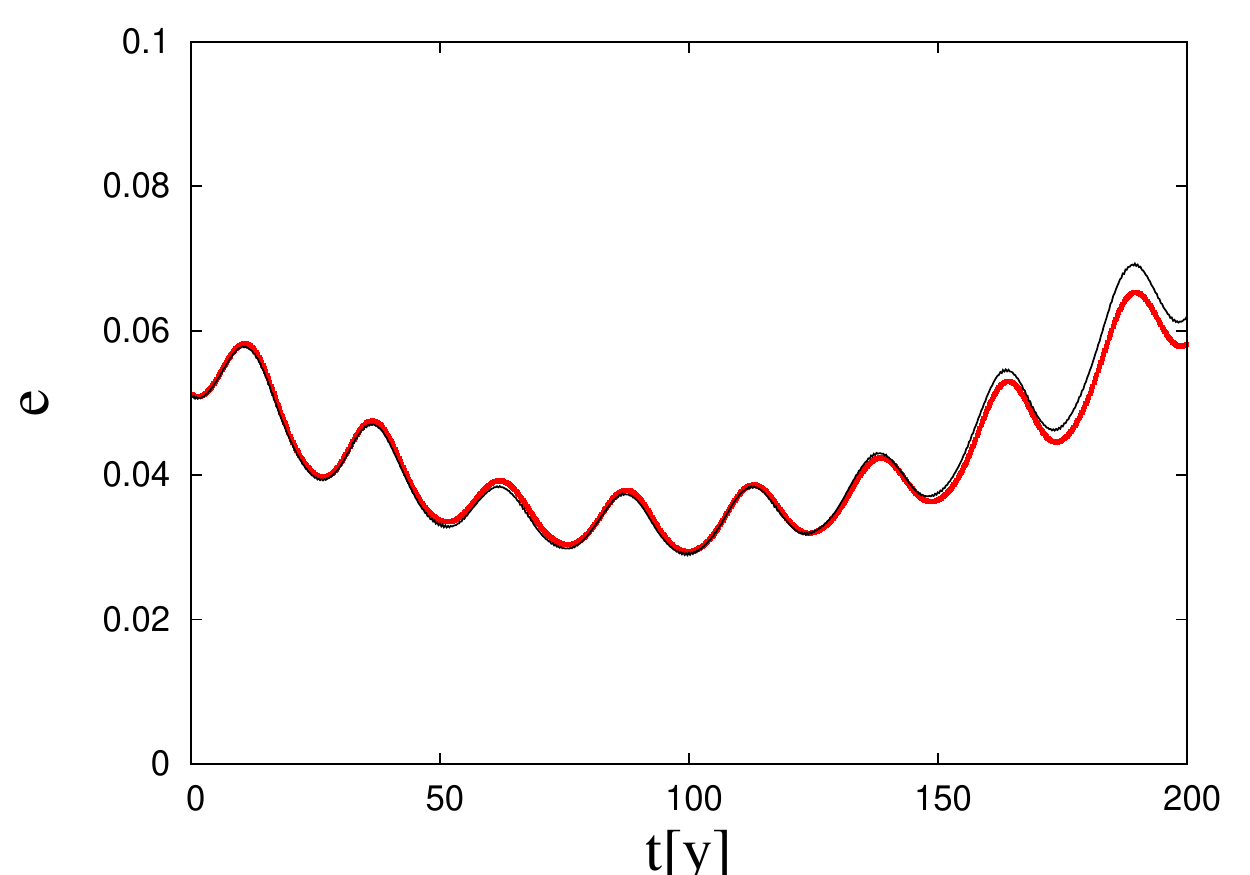}
\includegraphics[width=5truecm,height=4truecm]{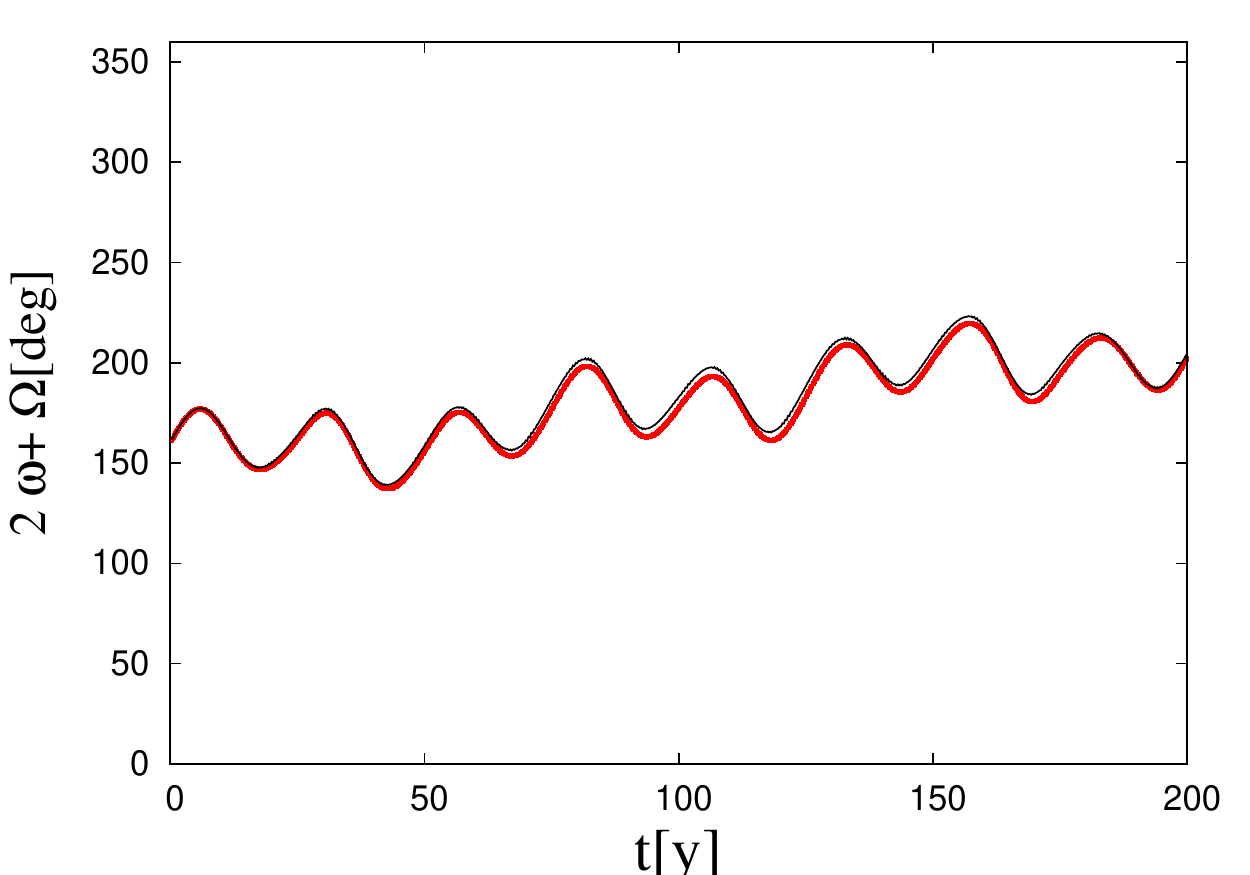}
\includegraphics[width=5truecm,height=4truecm]{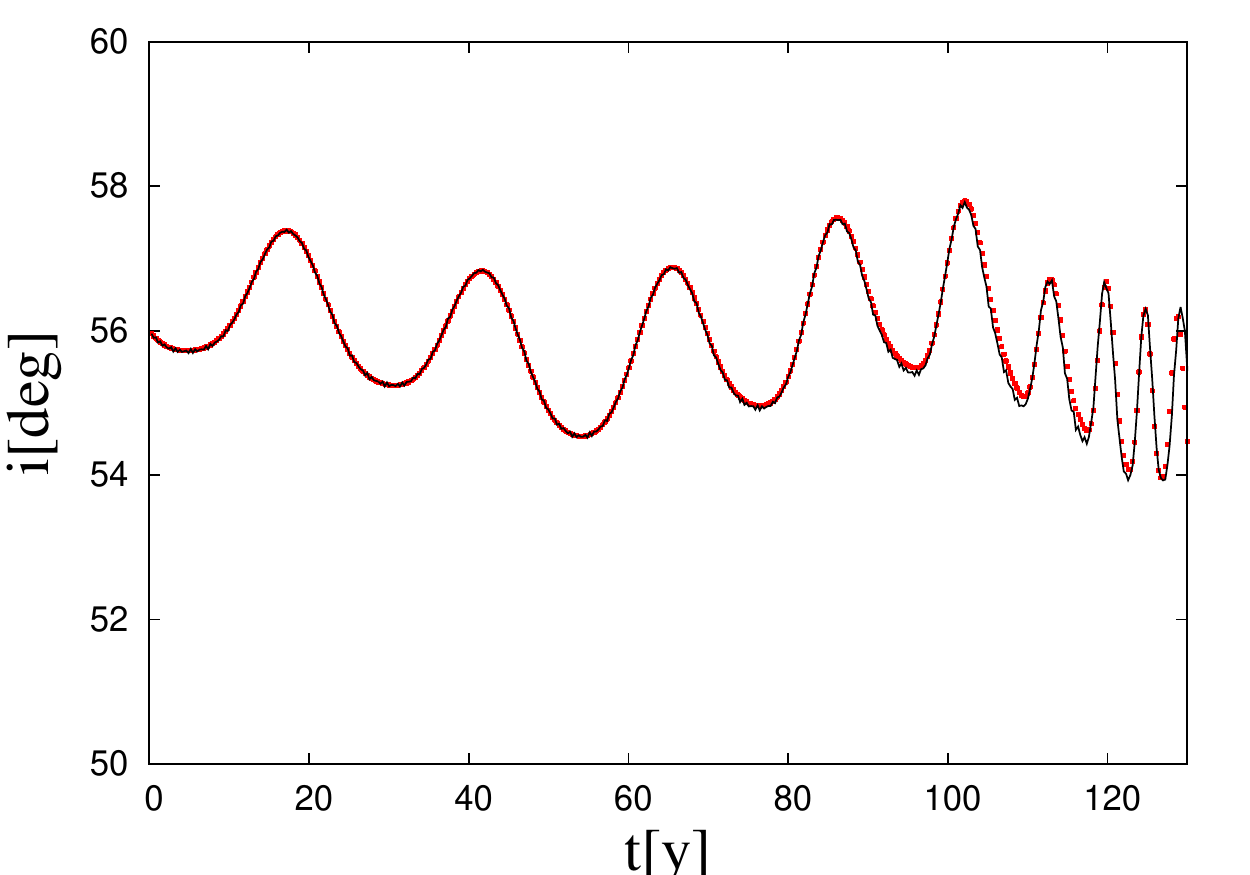}
\includegraphics[width=5truecm,height=4truecm]{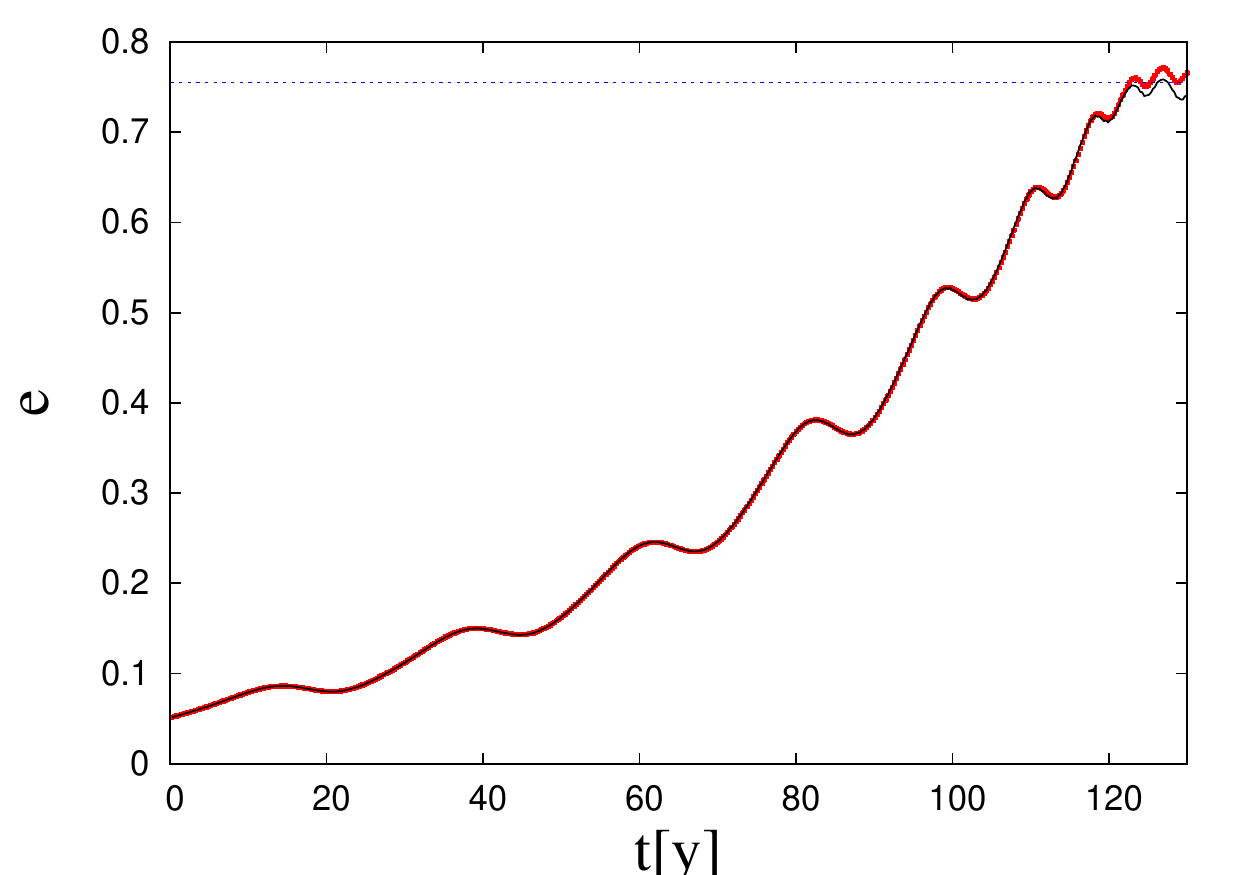}
\includegraphics[width=5truecm,height=4truecm]{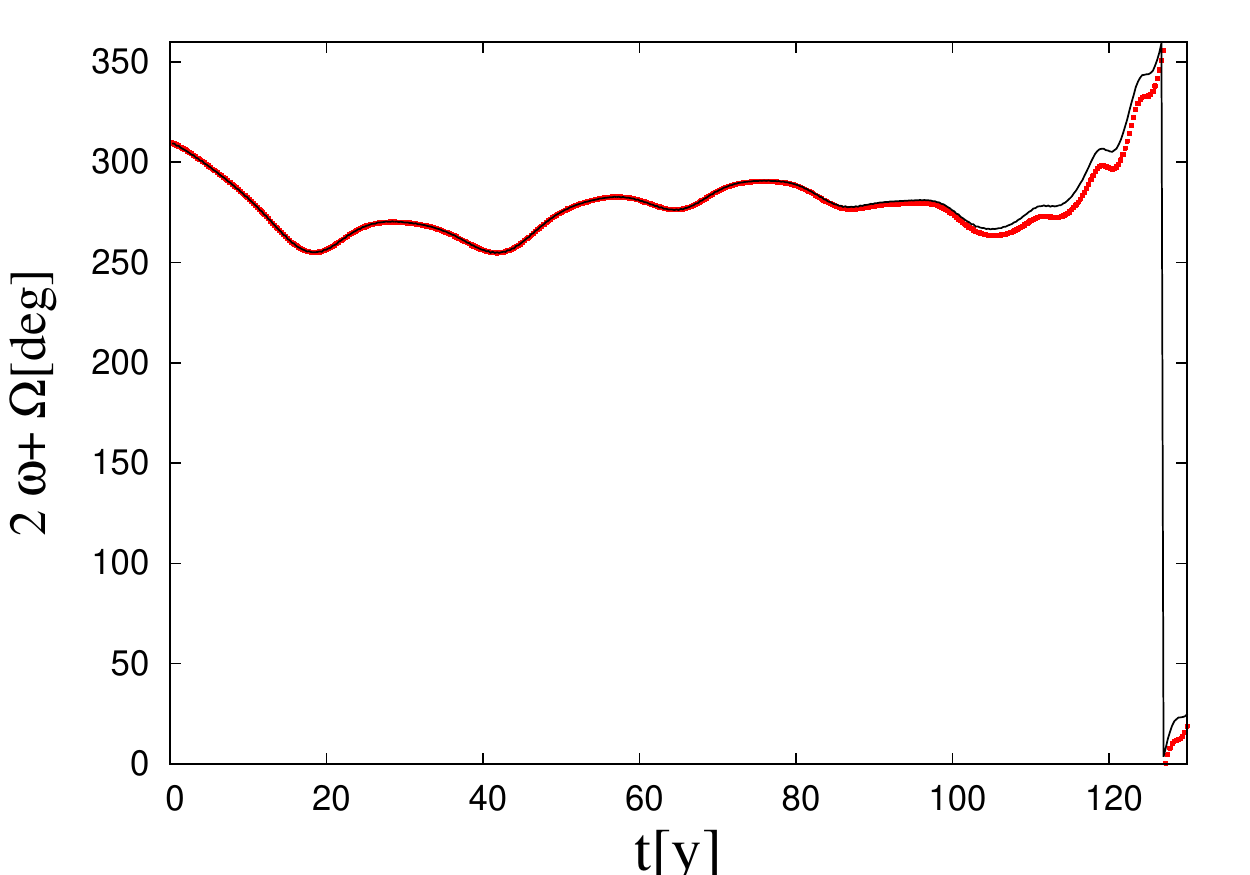}
\vglue0.5cm
\caption{Integration of the two orbits highlighted in the left panel of Figure~\ref{fig:lunisolar} (green-black circles).
The initial phase angles are $\omega(0)=190^o$ and $\Omega(0)=140^o$ for the top panels and,
respectively, $\omega(0)=30^o$ and $\Omega(0)=250^o$ for the bottom plots. The other data
are $a=26\,520\, km$,  $e(0)=0.05122$, $i(0)=56^o$. The results obtained by using the
Hamiltonian \eqref{ham_sec_res_all} are represented with the red color, while the black color
is used for the Cartesian model, which includes the Earth's gravity harmonics up to degree
and order $n=m=3$, the attraction of the Sun and Moon, as well as the influence of the solar radiation pressure
with $A/m=0.01$ $m^2/kg$. The horizontal line in the middle bottom plot indicates the eccentricity value leading to re-entry.
}
\label{fig:two_orbits}
\end{figure}

By shaping the long--term dynamics of satellites and space debris, lunisolar secular resonances play an essential
role in designing the end-of-life disposal strategies.
An extensive literature is devoted to the study of various disposal scenarios for the GNSS constellations,
by evaluating the effects induced by the secular resonances
(see \cite{ADRRVDM2016, chaogick, deleflie2011, RDAVRD, aR08, Sanchez15} and references therein).
It is beyond the scope of this paper to recall these strategies, but rather we focus on some dynamical features of the secular resonances. More precisely, we discuss some aspects concerning the eccentricity growth,
the overlapping of secular resonances, and the bifurcation of equilibria, by highlighting their effects on the long--term complex evolution of the medium Earth orbits.

Usually, in studying end-of-life disposal strategies, one investigates the eccentricity growth of the orbits located at several hundreds kilometers from the nominal constellation,  as a function of the initial phase angles $\Omega$ and $\omega$, and over a given interval of time of about
200 years (see \cite{ADRRVDM2016, RDGF2015, RDAVRD, Sanchez15}). The left panel of Figure~\ref{fig:lunisolar} is an eccentricity growth map, obtained by propagating the initial conditions $a=26\,520\, km$,  $e=0.05122$, $i=56^o$  for a large variety of initial orientation phases, and recording the maximum value of eccentricity
computed for each orbit. We stress that Figures~\ref{fig:lunisolar} and \ref{fig:two_orbits} may be considered as an application for the BeiDou constellation, since  we used the same value of the semimajor axis as that suggested in \cite{ADRRVDM2016} (see Table 2) for the eccentricity growth scenario.
However, our intention is not to discuss the possible disposal scenarios, but rather to interpret the eccentricity growth maps from the viewpoint of nonlinear dynamics.

The middle panel of Figure~\ref{fig:lunisolar} is obtained by considering $\Omega_M$ as a constant, while the other conditions are the same as for the
left plot of Figure~\ref{fig:lunisolar}.
Comparing the left and the middle panels of Figure~\ref{fig:lunisolar}, it is clear that the regression of the lunar nodes has a great influence in shaping the long-term dynamics of medium-Earth orbits. As it was pointed out in \cite{ElyHowell, aR15}, the motion of
the lunar nodes is responsible for the existence of a stochastic web with possible overlapping resonances, thus provoking the onset of chaos.

However, concerning the growth in eccentricity, the middle panel shows that this phenomenon could be explained as an effect of the single resonance  $2 \dot{\omega}+\dot{\Omega}=0$.
Along this line, we represent in the right plot of Figure~\ref{fig:lunisolar} the phase portrait of the one-degree-of-freedom toy model obtained
as follows. We consider the 2-dimensional, non-autonomous Hamiltonian \eqref{ham_sec_res_all};
after implementing a canonical transformation to pass to resonant variables, we average the Hamiltonian
over the non-resonant angle and then we take $\Omega_M$ as constant. The resulting Hamiltonian, to which we refer as
the \sl toy model, \rm has one degree of freedom (compare with \cite{CGfrontier, CGPbif}). Beside the eccentricity growth, this toy model shows another phenomenon, namely the bifurcation of equilibria. The right plot of Figure~\ref{fig:lunisolar} shows two
kinds of equilibrium points, the first one at high eccentricities (for $2\omega+\Omega = 360^o \,k$, $k\in \mathbb{Z}$),
and the other one at relatively small eccentricities (for $2\omega+\Omega = 180^o +360^o\,k$, $k\in \mathbb{Z}$). An initial condition taken inside the libration region corresponding to the equilibrium located at high eccentricities leads to a large excursion in eccentricity
and to a slow change of the resonant angle around $360^o\, k$, $k \in \mathbb{Z}$. On the contrary, inside the libration region corresponding to the other equilibrium point, one has smaller excursions in eccentricity, while the resonant angle varies around $180^o+360^o\, k$, $k \in \mathbb{Z}$.

The dynamical picture described above, predicted by an integrable toy model, could be considered as
starting point for understanding the evolution of real orbits. When the two degrees of freedom non--autonomous
Hamiltonian \eqref{ham_sec_res_all} is considered, then the complexity of the model increases;
for example, the phase space becomes four dimensional, the secular resonances can overlap,
the orbital elements may vary chaotically, the libration regions occupy different zones of the phase space.
However, by analysing the eccentricity growth map shown in Figure~\ref{fig:lunisolar}, we can identify the two kinds of
libration regions predicted by the toy model. Figure~\ref{fig:two_orbits} shows the evolution of two sample orbits,
represented by the green--black
circles in Figure~\ref{fig:lunisolar}. The orbit analysed in the top panels of Figure~\ref{fig:two_orbits} is located
inside the libration region that leads to small variations of the eccentricity and oscillations of the resonant angle around $180^o$. The other orbit is inside the libration region which leads to large excursions in eccentricity, while the resonant angle varies around $360^o$.  As it was stressed above, the variation of orbital elements is not regular due to the overlapping of resonances.

As a final remark, we mention that the results obtained by using the Hamiltonian \eqref{ham_sec_res_all} are validated in Figure~\ref{fig:two_orbits} by a comparison with the model developed in Cartesian
coordinates (see Section~\ref{sec:model}) that includes the geopotential, the gravitational attraction of Sun and Moon and the solar radiation pressure.

\section{Dissipative effects: the atmospheric drag}\label{sec:drag}
Above 50 $km$ from the Earth's surface the density of the atmosphere can be assumed sufficiently low to be approximated
as laminar air currents. Assuming that the atmosphere corotates with the Earth (i.e., disregarding the effect of winds),
neglecting the thermal motion of molecules and considering just accelerations in the direction of the satellite velocity
vector, the acceleration of the satellite due to atmospheric drag can be modeled as
$$
\mathbf{a}_d=-{C_D\over 2}\rho(|\mathbf{r}|){A\over m}\dot r'^2\ {{\dot {\mathbf{r}}'}\over {\dot{r}'}}\ ,
$$
where $\dot{\mathbf{r}}'$ is the velocity of the satellite relative to the particles with norm $\dot r'$, say
$$
\dot{\mathbf{ r}}'=\dot{\mathbf{ r}}-\boldsymbol{\omega}_E\wedge \mathbf{r}
$$
with $\boldsymbol{\omega}_E=\omega_E \mathbf{e}_3$ the angular velocity of the Earth
and where the coefficient $C_D$ can be assumed within $2\leq C_D\leq 2.5$, where $C_D=2.2$
holds for spherical satellites. Notice that
\beqano
\boldsymbol{\omega}_E\wedge \mathbf{r}&=& -\omega_E y \mathbf{e}_1+ \omega_E x \mathbf{e}_2 \,,\nonumber\\
\dot r'^2&=&(\dot{\mathbf{ r}}-\boldsymbol{\omega}_E\wedge \mathbf{r})\cdot (\dot{\mathbf{ r}}-\boldsymbol{\omega}_E\wedge \mathbf{r})=|\dot{\mathbf{ r}}|^2-2(\boldsymbol{\omega}_E\wedge \mathbf{r})\cdot \dot{\mathbf{ r}}+|\boldsymbol{\omega}_E\wedge \mathbf{r}|^2\nonumber\\
&=&(\dot x^2+\dot y^2+\dot z^2)-2\omega_E(x\dot y-\dot x y)+\omega_E^2(x^2+y^2)\ .
\eeqano
The density can be assumed to vary with the altitude above the surface, say $h=|\mathbf{r}|-R_E$; one can use the barometric formula:
$$
\rho(h)=\rho_0\ e^{-{{h-h_0}\over {H_0}}}\ ,
$$
where $\rho_0$ is the density at the reference altitude $h_0$ and $H_0$ is the scaling height at $h_0$.
Reference empirical values are given in Table~\ref{table:scaleheight} (\cite{LW}).

\vglue1cm
\begin{table}[h]
\begin{tabular}{|c|c|c|}
  \hline
  Altitude $h_0$ ($km$) & Atm. scale height $H_0$ ($km$) & Mean density $\rho_0$ ($kg/m^3$) \\
  \hline
  0 & 8.4 & $1.2$ \\
  200 & 37.5 & $2.53\cdot 10^{-10}$ \\
  400 & 58.2 & $2.72\cdot 10^{-12}$ \\
  600 & 74.8 & $1.04\cdot 10^{-13}$ \\
  800 & 151 &  $9.63\cdot 10^{-15}$ \\
  1000 & 296 & $2.78\cdot 10^{-15}$ \\
  1250 & 408 & $1.11\cdot 10^{-15}$ \\
  1500 & 516 & $5.21\cdot 10^{-16}$ \\
  2000 & 829 & $-$ \\
  \hline
 \end{tabular}
 \vskip.1in
 \caption{The atmospheric scale height and the mean density as a function of the altitude.}\label{table:scaleheight}
 \end{table}
\vglue1cm

In order to get the equations for $(L,G,H,M,\omega,\Omega)$ including the effect due to the atmospheric drag, we can
use the derivatives
\beqano
\dot L&=&{{\partial L}\over {\partial a}}\dot a\nonumber\\
\dot G&=&\dot L\sqrt{1-e^2}+{{\partial G}\over {\partial e}}\dot e\nonumber\\
\dot H&=&\dot G\cos i-G\sin i\ {{d i}\over {d t}}\ ,\nonumber\\
\eeqano
where, taking the average of the dissipative contribution over the mean anomaly (see \cite{chao}), we obtain:
\beqa{dissae}
\dot a&=&-{1\over {2\pi}}\int_0^{2\pi}B\rho v{a\over {1-e^2}}\ \Big[1+e^2+2e\cos f-\omega_E\cos i\sqrt{{a^3(1-e^2)^3}\over {\mu_E}}\Big]\ dM\ ,\nonumber\\
\dot e&=&-{1\over {2\pi}}\int_0^{2\pi}B\rho v\ \Big[e+\cos f-{{r^2\omega_E\cos i}\over{2\sqrt{\mu_E a(1-e)^2}}}\Big(2(e+\cos f)-e\sin^2f\Big)\Big]\ dM\ ,\nonumber\\
{{d i}\over {d t}} &= & 0\ ,
\eeqa
where $f$ is the true anomaly, $p=a(1-e^2)$, $\omega_E$ is the Earth's rotation rate, $\rho$ the atmospheric density, $B=C_D{A\over m}$ is
the ballistic coefficient and the satellite's velocity relative to the atmosphere is given by
$$
v=\sqrt{{\mu_E\over p}(1+e^2+2e\cos f)}\ \Big(1-{{(1-e^2)^{3\over 2}}\over {1+e^2+2e\cos f}}\, {\omega_E\over n^*}\cos i\Big)\ ,
$$
where $n^*$ is the mean motion of the satellite.

\begin{remark}
\begin{itemize}
    \item[$i)$] The equations \equ{dissae} are averaged over M and therefore $\dot M$ does not appear.
    \item[$ii)$] To simplify the computations of the averages, one can expand the arguments in the integrals in Fourier
    series of $M$.
    \item[$iii)$] The dissipative parameter is played by $B\rho$, where $C_D\sim O(1)$, $\rho\sim10^{-10}$ or even smaller  ($A\over m$
    is a conservative parameter). If $B=0$, then the system is conservative.
    \item[$iv)$] The dissipative effect influences just $\dot a$, $\dot e$ and not the other variables (in particular the angle variables).
   \end{itemize}

\end{remark}

\section{Minor effects}\label{sec:minor}
Beside the geopotential, lunar and solar attractions, and the solar radiation pressure, there exists a number of effects of less entity, which however should be considered when computing the long-term behavior of space debris. Below is a partial list,
which is definitely not exhaustive, but it might serve as an indication of some minor effects in space debris dynamics.

\begin{itemize}
    \item[$(1)$] Equinoctial precession of the Earth: due to this effect, the longitude of the equinox
    changes with respect to the ecliptic by $0.013845^o/yr$. Hence, the Earth-centered reference frame
    is non-inertial and this induces a long-period variation of the orbital elements, especially close
    to the stable and unstable equilibria (\cite{BelyaninGurfil}).
    \item[$(2)$] Earth's shadowing effects, which are taken into account at each revolution by computing
    the non-singular mean longitude at shadow entry and at shadow exit. The successive eclipses produce a
    fluctuation in the solar radiation caught by the debris (\cite{FerrazMello1972, HL2013, VL}).
    \item[$(3)$] Poynting-Robertson drag which, beside the solar radiation pressure, includes a dissipative
    contribution, whose effect is relevant on long time scales and for specific values of the orbital
    elements (\cite{LCG2016}).
    \item[$(4)$] Ocean tides, which induce a time variation of the spherical harmonic coefficients $C_{nm}$,
    $S_{nm}$ of the geopotential (see \cite{chao}).
    \item[$(5)$] Terrestrial tides, which provoke a time variation of the spherical harmonic coefficients
    as given in \cite{Peterson}, see also \cite{chao}.
    \item[$(6)$] Yarkovsky effect, which is a thermal force affecting the orbit of the debris.
    \item[$(7)$] Earth's radiation pressure emitted from the Earth with two components:
    the infrared and the optical radiation.
    \item[$(8)$] Relativistic effects, inducing a relativistic acceleration on the satellite,
    which is composed by several factors, like the spherical central body term and its oblateness correction,
    the geodesic precession, the relativistic rotational energy and the Lense-Thirring acceleration.
    \item[$(9)$] Planetary forces, which are relevant at high altitudes, especially in GEO and for external
    resonances.
\end{itemize}

\section{Conclusions}\label{sec:conclusions}
The awareness that space debris can produce serious concerns for Earth orbiting satellites has increased the interest toward
the dynamics of small objects orbiting our planet. The models describing the dynamics vary according to
the altitude of the space debris. For this reason it is essential to understand where the object
is located and which are the main forces acting on it. Once the model is defined, one can proceed to
investigate the dynamics using the appropriate formulation: Cartesian, Delaunay, Milankovitch,
epicyclic variables. A special role is played by resonances which are of different type:
tesseral, secular and semi-secular resonances. This classification leads to highlight
different factors which contribute to the onset of chaos as some parameters are varied.

The overlapping of tesseral resonances is a well-known source of chaos, which might be obtained by a change of
the orbital parameters, in particular the inclination and/or eccentricity, or rather by increasing the area-to-mass ratio of the space debris.
This result leads to design possible disposal strategies, which are reminiscent of the methods for
describing interplanetary trajectories by using low-energy orbits: whenever the
change of parameter generates chaos, one can move the space debris within different regions.
Another important role in designing disposal orbits is played by lunisolar resonances, especially
within the GNSS constellations. Remarkable effects due to such resonances are the growth of
the eccentricity or the appearance of bifurcations of equilibria, as it can be deduced from the analysis
of the averaged Hamiltonian.

The success of using analytical tools to justify unusual phenomena or to foresee the dynamics motivates
this work, which collects all major formalisms for the study of space debris dynamics and hopefully it
will serve as reference for future developments.

\begin{appendices}

\section*{Appendix: Chaos Indicators} \label{sec:chaosIndicators}

In order to investigate the stability of the dynamics, several tools have been introduced. Among them, the most familiar one is the
Lyapunov characteristic exponent (\cite{BGS}). However, for our particular needs, it is more convenient and useful to compute a quantity called the Fast Lyapunov Indicator,
hereafter FLI (\cite{froes}). Here we briefly recall how these quantities are defined.

The Lyapunov characteristic exponent provides evidence of the chaotic character of the dynamics of a given dynamical system, since it measures the divergence
of nearby trajectories. For a phase space of dimension $N$, there exist $N$ Lyapunov exponents, although the largest one is the most significative and is what we
refer to as the Lyapunov exponent. This is due to the exponential rate of divergence, so that the greatest exponent dominates the overall separation.

A practical procedure to compute the  Lyapunov exponents is the following (\cite{LAVW}): let ${\underline \xi}=(L,G,H,\ell M-j\theta,\omega,\Omega)$ be the phase state associated with
the Hamiltonian, e.g. \eqref{ham_tess_res_all}. We can generically denote the evolution in phase space as determined by the vector field
$$
\dot{\underline \xi}={\underline f}({\underline \xi})\ ,\qquad {\underline{\xi}}\in \mathbb{R}^6\ ,
$$
and the evolution on the tangent space by the corresponding variational equations
$$
\dot{\underline \eta}=\left({{\partial \underline{f}(\underline{\xi})} \over {\partial
\underline{\xi}}}\right)\ {\underline \eta}\ ,\qquad {\underline{\eta}}\in \mathbb{R}^6\ .
$$
We can assign the initial conditions by choosing $\underline{\xi}(0)$ and
each component of $\underline{\eta}(0) = {\eta}_j  (0) \, {\hat e}_j$ in a basis ${\hat e}_j$ of the tangent space. Then,
we can compute the quantities
$$
\chi_j \equiv \lim_{t\to\infty}  \lim_{|| \underline{\eta} (0)|| \to 0}{1\over t}
\log {{|{\eta}_j (t)|}\over {|{\eta}_j (0)|}}\ , \qquad j=1,...,6 \ ,
$$
where $\|\cdot\|$ denotes the Euclidean norm.
When dealing with a {\it Hamiltonian dynamical system} only $N/2$ of the $\chi_j$ are actually meaningful, so in our case we would have three exponents. In view of the exponential rate of divergence, we can concentrate on the greatest of them and estimate it by means of the formula
$$
\chi \equiv \lim_{t\to\infty} {1\over {t}} \log {{|| \underline{\eta} (t)||}\over {|| \underline{\eta} (0)||}}\ ,
$$
where $|| \underline{\eta} (t)||$ is the phase-space distance at time $t$ between
trajectories at initial distance $|| \underline{\eta} (0)||$. In practice, to overcome overflows, this procedure is implemented by dividing the whole time-span $t$ into a set of sampling times $\tau$ and renormalizing the solution $|| \underline{\eta} (n \tau)||, n=1,2,...$ at each sampling
time (\cite{BGS}).


In order to investigate the stability of the dynamics for the models described
in the previous sections, we prefer to compute the Fast Lyapunov Indicator, which is defined as the value of the largest Lyapunov characteristic exponent
{\it at a fixed time} (see \cite{froes}). By comparing the values of the FLIs as initial conditions or parameters are varied, one obtains
an indication of the dynamical character of the phase-space trajectories as well as of their chaoticity/regularity behaviour.
The explicit computation of the FLI proceeds as follows: the FLI at a given time $T\geq 0$ is obtained by the expression
$$
{\rm FLI}(\underline{\xi}(0),  \underline{\eta}(0), T) \equiv \sup _{0 < t\leq T}
 \log || \underline{\eta}(t)||\ .
$$
In practice, a \sl reasonable \rm choice of $T$ makes faster the computation of the FLI when compared with
previous expressions for the $\chi$'s where, in principle, very long integration times are required
to obtain a reliable convergence process.

\end{appendices}

\section*{Acknowledgements}
A.C. and G.P. were partially supported by the Stardust Marie Curie Initial Training Network, FP7-PEOPLE-2012-ITN, Grant
Agreement 317185 and by GNFM/INdAM.
F.G. was supported by the Stardust Marie Curie Initial Training Network, FP7-PEOPLE-2012-ITN, Grant Agreement 317185. C.G. was supported by a grant of the Romanian National Authority for
Scientific Research and Innovation, CNCS - UEFISCDI, project number
PN-III-P4-ID-PCE-2016-0182.


\begin{thebibliography}{9}


\bibitem{ADRRVDM2016}
E.M. Alessi, F. Deleflie, A.J. Rosengren
A. Rossi, G.B. Valsecchi, J. Daquin, K. Merz, \sl
A numerical investigation on the eccentricity growth
of GNSS disposal orbits, \rm  Celest. Mech. Dyn. Astr.  {\bf 125}, 71--90 (2016)

\bibitem{AC}
R.R. Allan, G.E. Cook,
\sl The long period motion of the plane of a distant circular orbit, \rm
RSPSA {\bf 280}, n. 1380, 97-109 (1964)

\bibitem{VS}
P.V. Anderson, H. Schaub,
\sl Local orbital debris flux study in the geostationary ring, \rm
Adv. Space Res. {\bf 51}, 2195-2206 (2013)

\bibitem{LAVW}
L. Arnold,  V. Wihstutz, \sl Lyapunov exponents, a Survey, \rm Lecture Notes in Mathematics, {\bf 1186}, 1--26, Springer (1986)

\bibitem{BelyaninGurfil}
S. Belyanin, P. Gurfil, \sl Semianalytical study of geosynchronous orbits about a precessing oblate Earth under lunisolar gravitation and tesseral resonance, \rm Journal of the Astronautical Sciences {\bf 57}, 513--543 (2010)

\bibitem{BGS}
G. Benettin, L. Galgani, A. Giorgilli, J.-M. Strelcyn, \sl Lyapunov characteristic exponents for
smooth dynamical systems and for Hamiltonian systems: A method for computing all of them, \rm
Meccanica {\bf 15}, 9-–30 (1980)

\bibitem{Breiter2001}
S. Breiter, \sl Lunisolar resonances revisited, \rm Celest. Mech. Dyn. Astron.
{\bf 81}, 81--91 (2001)

\bibitem{BWM}
S. Breiter, I. Wytrzyszczak, B. Melendo, \sl Long--term predictability of
orbits around the geosynchronous altitude, \rm Adv. Space Res. {\bf 35}, 1313--1317, (2005)

\bibitem{CPL2015}
D. Casanova, A. Petit, A. Lema\^{\i}tre, \sl
Long-term evolution of space debris under the $J_2$ effect, the solar radiation pressure and the solar and lunar perturbations, \rm
Celest. Mech. Dyn. Astr. {\bf 123}, 223--238 (2015)

\bibitem{CGmajor}
A. Celletti, C. Gale\c s, \sl On the dynamics of space debris: 1:1 and 2:1 resonances, \rm
J. Nonlinear Science {\bf 24}, 1231--1262 (2014)

\bibitem{CGminor}
A. Celletti, C. Gale\c s, \sl Dynamical investigation of minor resonances for space debris, \rm Celest. Mech. Dyn. Astr.
{\bf 123},  203-222 (2015)

\bibitem{CGext}
A. Celletti, C. Gale\c s, \sl A study of the main resonances outside the geostationary ring, \rm Advan. Space Res.
{\bf 56}, 388--405 (2015)

\bibitem{CGleo}
A. Celletti, C. Gale\c s, \sl Dynamics of resonances and equilibria of low Earth objects, \rm Preprint (2016)

\bibitem{CGfrontier}
A. Celletti, C. Gale\c s, \sl
A study of the lunisolar secular resonance $2\dot{\Omega}+\dot{\omega}=0$, \rm Frontiers in Astronomy and Space Sciences, {\bf 3}, 11 pages (2016)

\bibitem{CGPbif}
A. Celletti, C.  Gale\c s, G. Pucacco, {\sl Bifurcation of lunisolar secular resonances for space debris orbits},
SIAM J. Appl. Dyn. Syst. {\bf 15}, 1352--1383 (2016)

\bibitem{CGPRnote}
A. Celletti, C. Gale\c s, G. Pucacco, A. Rosengren, \sl Analytical development of the lunisolar disturbing function and the critical
inclination secular resonance, \rm Celest. Mech. Dyn. Astr., DOI: 10.1007/s10569-016-9726-8 (2016)

\bibitem{chao}
C.G. Chao, {\sl Applied orbit perturbation and maintenance}, Aerospace Press Series, AIAA, Reston, Virgina (2005)

\bibitem{chaogick}
C.C. Chao, R.A. Gick, \sl
Long-term evolution of navigation satellite
orbits: GPS/GLONASS/GALILEO,\rm \, Adv. Space Res. {\bf 34}, 1221--1226 (2004)

\bibitem{cook1962}
G.E. Cook, \sl Luni-Solar perturbations of the orbit of an Earth satellite, \rm  Geophys. J. R. astr. Soc. {\bf  6},
271--291 (1962)

\bibitem{DRADVR15}
J. Daquin, A.J. Rosengren, E.M. Alessi, F. Deleflie, G.B. Valsecchi, A. Rossi, \sl The dynamical structure of the MEO region: long-term
stability, chaos, and transport, \rm Celest. Mech. Dyn. Astr. {\bf 124}, 335--366 (2016)

\bibitem{deleflie2011}
F. Deleflie, A. Rossi, C. Portmann, G. Me´tris, F. Barlier, \sl
Semi-analytical investigations of the long term evolution of the eccentricity of Galileo and GPS-like orbits, \rm Adv. Space Res. {\bf 47}, 811--821 (2011)

\bibitem{YX2}
A. Deprit, \sl Canonical transformations depending on a small parameter, \rm
Celest. Mech. {\bf 1}, n. 1, 12-30 (1969)


\bibitem{YX3}
C. Efthymiopoulos, \sl Canonical perturbation theory; stability and diffusion
in Hamiltonian systems: applications in dynamical astronomy, \rm in P.M. Cincotta,
C.M. Giordano, C. Efthymiopoulos, eds., ``Third La Plata International School on
Astronomy and Geophysics", Workshop Series of the Asociacion
Argentina de Astronomia {\bf 3}, 3-146 (2012)


\bibitem{EGM2008} Earth Gravitational Model 2008, $http://earth-info.nga.mil/GandG/wgs84/gravitymod/egm2008/$

\bibitem{ElyHowell}
T.A. Ely, K.C. Howell, {\sl Dynamics of artificial satellite orbits with tesseral resonances including the effects of luni--solar
perturbations}, Dynamics and Stability of Systems {\bf 12},  243--269 (1997)

\bibitem{FerrazMello1972}
S. Ferraz Mello, \sl
Analytical study of the Earth's shadowing effects on satellite orbits, \rm
Celest. Mech. Dyn. Astron. {\bf 5},  80--101 (1972)

\bibitem{FKZ}
L.J. Friesen, D.J. Kessler, H.A. Zook,
\sl Reduced debris hazard resulting from a stable inclined geo-synchronous orbit, \rm
Adv. Space Res. {\bf 13}, n. 8, 231-241 (1993)

\bibitem{froes}
C. Froeschl\'e, E. Lega, R. Gonczi, \sl Fast Lyapunov indicators. Application to asteroidal motion, \rm Celest. Mech. Dyn. Astr. {\bf 67}, 41--62 (1997)

\bibitem{Gachet2016}
F. Gachet, A. Celletti, G. Pucacco, C. Efthymiopoulos, \sl
Geostationary secular dynamics revisited:
application to high area-to-mass ratio objects, \rm to be
published in Celest. Mech. Dyn. Astr. (2017)

\bibitem{GDGR}
I. Gkolias, J. Daquin, F. Gachet, A.J. Rosengren, \sl From order to chaos
in Earth satellite orbits, \rm The Astronomical Journal {\bf 152}, n. 5,
119 (2016)


\bibitem{gG74}
G.E.O. Giacaglia, {\sl Lunar perturbations of artificial satellites of the Earth},
Celest. Mech. {\bf 9}, 239--267 (1974)

\bibitem{giacaglia} G.E.O. Giacaglia, \sl A note on Hansen's coefficients in satellite theory, \rm
Celest. Mech. {\bf 14}, 515-523 (1976)

\bibitem{YX1}
G.-I. Hori, \sl Theory of general perturbation with unspecified canonical
variable, \rm Publ. Astron. Soc. Japan {\bf 18}, n. 4, 287-296 (1966)


\bibitem{HL2013}
Ch. Hubaux, A. Lema\^{\i}tre, \sl The impact of Earth's shadow on the long-term evolution of space debris, \rm
Celest. Mech. Dyn. Astr. {\bf 116}, 79-95 (2013)

\bibitem{HughesI}
S. Hughes, {\sl Earth satellite orbits with resonant lunisolar perturbations. I. Resonances dependent only on inclination},
Proc. R. Soc. Lond. A {\bf 372}, 243--264 (1980)

\bibitem{HughesII}
S. Hughes, {\sl Earth satellite orbits with resonant lunisolar pertubations. II. Some resonances dependent on the semi-major axis, eccentricity and inclination}, Proc. R. Soc. Lond. A {\bf 375}, 379--396 (1981)

\bibitem{IADC}
IADC space debris mitigation guidelines,
$http://www.iadc-online.org/index.cgi?item=docs\_pub$

\bibitem{wK62}
W. M. Kaula,
{\sl Development of the lunar and solar disturbing functions for a close satellite},
Astron. J. {\bf 67}, 300--303 (1962)

\bibitem{wK66}
W. M. Kaula, {\sl Theory of satellite Geodesy},
Blaisdell, Waltham (1966)

\bibitem{X1}
D.J. Kessler, B.G. Cour-Palais, \sl Collision frequency of artificial
satellites: The creation of a debris belt, \rm J. Geophys. Res. {\bf 83}, n. A6,
2637-2646 (1978)


\bibitem{KH1958}
D.G. King-Hele, {\sl The effect of the Earth's oblateness on the orbit of a near satellite},  Proc. R. Soc. Lond. A {\bf 247}, 49--72 (1958)

\bibitem{X2}
H. Klinkrad, \sl Space Debris, \rm Springer Praxis Books, Springer Berlin
Heidelberg (2006)


\bibitem{Kozai1959}
Y. Kozai, \sl On the effects of the Sun and the Moon upon the motion
of a close Earth satellite, \rm Smithsonian Astrophys. Observ. Spec.
Rept. 22: 7-10, Cambridge, Mass. March 20, (1959)

\bibitem{Kozai66}
Y. Kozai, \sl Lunar perturbations with short periods, \rm
Smithsonian Astrophys. Obs. Spec. Rep. \textbf{235} (1966)

\bibitem{KH}
P.H. Krisko, D.T. Hall,
\sl Geosynchronous region orbital debris modeling with GEO$\_$EVOLVE 2.0, \rm
Adv. Space Res. {\bf 34} 1166-1170 (2004)

\bibitem{JAH}
R. Jehn, V. Agapov, C. Hernadez,  
\sl End-of-Life disposal of Geostationary satellites, \rm Proc. Fourth European Conference on Space Debris
ESA SP587 {\bf373J} (2005)

\bibitem{J2012}
N.L. Johnson,
\sl A new look at the GEO and near-GEO regimes: Operations, disposals, and debris, \rm
Acta Astronautica {\bf 80} 82-88 (2012)

\bibitem{mL89}
M. T. Lane,
{\sl On analytic modeling of lunar perturbations of artificial satellites of the Earth},
Celest. Mech. Dyn. Astr. {\bf 46}, 287--305 (1989)

\bibitem{LW}
W. Larson, J. Wertz, \sl Space mission analysis and design, \rm Kluwer publ. (1999)

\bibitem{LDV}
A. Lema\^{\i}tre, N. Delsate, S. Valk,
\sl A web of secondary resonances for large $A/m$ geostationary debris, \rm Celest. Mech. Dyn. Astr. {\bf 104}, 383-402 (2009)

\bibitem{XX2}
J.-C. Liou, J.K. Weaver, \sl Orbital Dynamics of High Area-To-Mass Ratio
Debris and Their Distribution in the Geosynchronous Region, \rm in D. Danesy,
editor, 4th European Conference on Space Debris {\bf 587}, ESA Special
Publication, 285-290 (2005)


\bibitem{LCG2016}
C. Lhotka, A. Celletti, C. Gale\c s, \sl Poynting-Robertson drag and solar wind in the space debris problem, \rm Mon. Not. R. Astron. Soc. {\bf 460}, 802--815 (2016)

\bibitem{Milankovitch1939}
M. Milankovitch, {\sl \"{U}ber die Verwendung vektorieller Bahnelemente in der St\"{o}rungsrechnung,} Bull. Acad. Sci. Math. Nat. A \textbf{6}, 1--70 (1939)

\bibitem{MG}
O. Montenbruck, E. Gill, \sl Satellite orbits, \rm Science, {\bf 134}, Springer-Berlin-Heidelberg (2000)

\bibitem{cMsD99}
C.D. Murray, S.F. Dermott,
{\sl Solar System Dynamics},
Cambridge University Press, Cambridge (1999)

\bibitem{musen1961}
P. Musen, A. Bailie,  and E. Upton, \sl Development of the lunar
and solar perturbations in the motion of an artificial satellite, \rm
NASA Technical Note D--494, p. 40 (1961)

\bibitem{PA}
C. Pardini, L. Anselmo,
\sl Long-Term Evolution of Geosynchronous Orbital Debris with High Area-to-Mass Ratios, \rm
Trans. Japan Soc. Aero. Space Sci. {\bf 51}, n. 171, 22-27 (2008)

\bibitem{Peterson}
G.E. Peterson, \sl Estimation of the Lense-Thirring precession using Laser--ranged satellites, \rm Ph.D. thesis, University of Texas at Austin (1997)

\bibitem{RDGF2015}
J. Radtke, R. Dominguez-Gonzalez, S.K.  Flegel,  N.  Sanchez-Ortiz, K.  Merz, \sl Impact of eccentricity build-up and graveyard disposal strategies on MEO navigation constellations, \rm  Adv. Space Res.
\textbf{56}, 2626--2644 (2015)

\bibitem{aR15}
A.J. Rosengren, E.M. Alessi, A. Rossi, G.B. Valsecchi,
{\sl Chaos in navigation satellite orbits caused by the perturbed motion of the Moon},
Mon. Not. R. Astron. Soc. {\bf 449}, 3522--3526 (2015)

\bibitem{RDAVRD}
A.J. Rosengren, J. Daquin, E.M. Alessi, G.B. Valsecchi, A. Rossi,  F.
Deleflie, \sl GALILEO Disposal Orbit Strategy: Resonances, Chaos and
Stability, \rm Mon. Not. R. Astron. Soc.  {\bf 464}, n. 4, 4063-4076 (2017)


\bibitem{Rosengren2013}
A.J. Rosengren, D.J. Scheeres, {\sl Long-term dynamics of high area-to-mass ratio objects in high-Earth orbit,}
Adv. Space Res. \textbf{52}, 1545--1560 (2013)

\bibitem{Rosengren2014}
A.J. Rosengren, D.J. Scheeres,  {\sl On the Milankovitch orbital elements for perturbed Keplerian motion,} Celest. Mech. Dyn. Astr., \textbf{118}, 197--220 (2014)

\bibitem{RS}
A.J. Rosengren, D.J. Scheeres,
\sl Laplace Plane modifications arising from solar radiation pressure, \rm
The Astroph. J. {\bf 786}, n. 45, pp. 13 (2014)

\bibitem{aR08}
A. Rossi, \sl  Resonant dynamics of Medium Earth Orbits: space debris issues, \rm Celest. Mech. Dyn. Astr. {\bf 100}, 267--286 (2008)

\bibitem{Sanchez15}
D. M. Sanchez, T.  Yokoyama, A.F.B. de Almeida Prado, \sl Study of some strategies for disposal of the GNSS satellites, \rm Mathematical Problems
in Engineering, Volume {\bf 2015}, Article ID 382340, 14 pages (2015)

\bibitem{XX1}
T. Schildknecht, \sl  Optical surveys for space debris, \rm The Astronomy and
Astrophysics Review {\bf 14}, n. 1, 41-111 (2007)


\bibitem{Simon1994}
J.L. Simon, P. Bretagnon, J. Chapront, M, Chapront--Touz\'e, G. Francou, J. Laskar, \sl Numerical expressions for precession formulae and mean elements for the Moon and the planets, \rm Astron. Astrophys. {\bf 282}, 663--683 (1994)

\bibitem{Tremaine2009}
S. Tremaine, J. Touma, F. Namouni, {\sl Satellite dynamics on the Laplace surface,} The Astronomical Journal, \textbf{137}, 3706--3717 (2009)

\bibitem{upton1959}
E. Upton, A. Bailie, P. Musen, \sl Lunar and solar perturbations on satellite orbits, \rm
Science {\bf 130}, n. 3390, 1710-1 (1959)

\bibitem{VLA}
S. Valk, A. Lema\^{\i}tre, L. Anselmo, \sl Analytical and semi-analytical investigations of geosynchronous space debris with high area-to-mass ratios, \rm Adv. Space Res. {\bf 41}, 1077-1090 (2008)

\bibitem{VL}
S. Valk, A. Lema\^{\i}tre, \sl Semi-analytical investigations of high area-to-mass ratio geosynchronous space debris including Earth's shadowing effects, \rm Adv. Space Res. {\bf 42}, 1429-1443 (2008)

\bibitem{VLD}
S. Valk, A. Lema\^{\i}tre, F. Deleflie, \sl Semi-analytical theory of mean orbital motion for geosynchronous space debris under gravitational influence, \rm Adv. Space Res. {\bf 43}, 1070-1082 (2009)

\bibitem{VDLC}
S. Valk, N. Delsate, A. Lema\^{\i}tre, T. Carletti, \sl Global dynamics of high area-to-mass ratios
geosynchronous space debris by means of the MEGNO indicator, \rm Adv. Space Res. {\bf 43}, 1509--1526 (2009)

\bibitem{zhuetal2015}
T.L. Zhu, C.Y. Zhao, H.B. Wang, M.J. Zhang,
\sl Analysis on the long term orbital evolution of Molniya satellites, \rm
Astrophys Space Sci. {\bf 357}, 126 (2015)





\end{thebibliography}
\end{document}